%% file: CRdust22.tex
\documentclass[manuscript]{aastex}

\usepackage{adjustbox}

\slugcomment{Submitted to The Astrophysical Journal Supplement Series}

\shorttitle{Interstellar grain heating by cosmic rays}
\shortauthors{Kalv\=ans}

\begin{document}

\title{Temperature spectra of interstellar dust grains heated by cosmic-rays I:\\ translucent clouds}

\author{Juris Kalv\=ans}
\affil{Engineering Research Institute ``Ventspils International Radio Astronomy Center" of Ventspils University College,\\
Inzenieru~101, Ventspils, LV-3601, Latvia}
\email{juris.kalvans@venta.lv}

\begin{abstract}
Heating of whole interstellar dust grains by cosmic-ray (CR) particles affects the gas-grain chemistry in molecular clouds by promoting molecule desorption, diffusion, and chemical reactions on grain surfaces. The frequency of such heating $f_T$, s$^{-1}$, determines how often a certain temperature $T_{\rm CR}$,~K, is reached for grains hit by CR particles. This study aims to provide astrochemists with comprehensive and updated dataset on the CR-induced whole-grain heating. We present calculations of $f_T$ and $T_{\rm CR}$ spectra for bare olivine grains with radius $a$ of 0.05; 0.1; 0.2~$\mu$m, and such grains covered with ice mantles of thickness 0.1$a$ and 0.3$a$. Grain shape and structure effects are considered, as well as 30 CR elemental constituents with an updated energy spectrum corresponding to a translucent cloud with $A_V=2$~mag. Energy deposition by CRs in grain material was calculated with the \textsc{srim} program. We report full $T_{\rm CR}$ spectra for all nine grain types and consider initial grain temperatures of 10~K and 20~K. We also provide frequencies for a range of minimum $T_{\rm CR}$ values. The calculated dataset can be simply and flexibly implemented in astrochemical models. The results show that, in the case of translucent clouds, the currently adopted rate for heating of whole grains to temperatures in excess of 70~K is underestimated by approximately two orders of magnitude in astrochemical numerical simulations. Additionally, grains are heated by CRs to modest temperatures (20--30~K) with intervals of a few years, which reduces the possibility of ice chemical explosions.
\end{abstract}

\keywords{ISM:dust --- cosmic rays --- astrochemistry}

\section{Introduction}
\label{intro}

Among the various processes that affect dust grains in the interstellar medium is interaction with cosmic-ray (CR) particles. These may deposit a considerable amount of energy into the grains, which induces whole-grain heating. Studies of such events have been inspired by their possible effect on the gas-grain chemistry in dark cloud cores and star-forming regions. The initial research by \citet*{Leger85} studied the effects of olivine grain collisions with iron nuclei CRs, while those of \citet{Shen04,deBarros11b,deBarros11d} and \citet{Ivlev15r} covered the deposition of CR energy into grains in additional detail. Grain heating events are of interest because they serve as an energy source for some surface processes in dark clouds. Such processes include chemical explosions \citep{Leger85,Shen04}, molecule evaporation \citep{Hasegawa93,Willacy93,Roberts07}, diffusion \citep{Reboussin14,Kalvans14ba}, and chemical reactions \citep{Kalvans15a} for molecules adsorbed on interstellar or circumstellar grains.

The above studies investigate molecular processes, approximating CR-induced grain heating to a minimum temperature $T_{\rm CR}$, K, threshold most relevant for the process in consideration. Each $T_{\rm CR}$ threshold has its characteristic occurrence frequency $\Sigma f_T$, s$^{-1}$. For example, a 70~K threshold was adopted by \citet{Hasegawa93} to describe evaporation, while 27~K was adopted by \citet{Leger85} as a sufficient temperature to enable radical mobility on grain surfaces, triggering chemical explosions in the ice layer.

The interstellar grains are affected by CR particles consisting of many elements with various energies and frequencies. Additionally, the CR hits deposit different proportions of their energy into the grains, depending on the properties of the particular CR particle, its trajectory through the grain, and grain material. The situation becomes more complex, when inhomogeneous grains are considered. The resulting CR-induced heating can be described with a spectrum of grain temperatures $T_{\rm CR}$, each of which occurs with a certain frequency $f_T$.

The aim of this paper is to present a dedicated study to compute such a spectrum. This involves two tasks. The first is to calculate the data considering as much detail as necessary, within reason. The second task is to present the results in a comprehensive but convenient manner. To fulfil this, we include the whole temperature spectrum for CR-heated grains, not just a few threshold temperatures and their occurrence frequencies, as in previous studies. Such a $T_{\rm CR}$ spectrum will allow, for example, to include whole-grain heating in astrochemical models that study certain physical or chemical processes (evaporation, surface diffusion, reactions) that require a characteristic minimum grain temperature. The study was motivated by the lack of published proper whole-grain heating data as discussed by \citet{Kalvans15a}.

It is important to clearly state that we do not consider the cooling of the grains. Cooling determines how long a heated grain will remain within a certain temperature range. Cooling by evaporation of volatiles, such as the CO molecule, is likely to be the main mechanism in molecular clouds \citep{Leger85}. The number of evaporated CO molecules \citep[$>10^5$,][]{Roberts07} is comparable to or lower than the number of adsorption sites on grain surface. This means that the existence of even a very thin ice layer may make evaporative cooling the dominant mechanism, which means that the cooling rate depends on the chemical composition of the ice layer. The computation of ice composition and thus, an accurate cooling rate, is a task for astrochemical simulations. Here, we deal exclusively with grain heating by CRs to a certain temperature $T_{\rm CR}$, which is reached before cooling processes become significant.

\section{Methodology}
\label{meth}

A number of aspects have to be taken into account when calculating the temperature spectrum of CR-heated grains. Cosmic rays of different energies $E_{\rm CR}$ (MeV/ion or MeV/amu) have different abundances, i.e., they have a characteristic energy spectrum. This initial spectrum is modified by the medium, where the CR particles propagate. The CR particles include many chemical elements, each of which also has a different abundance.

Various cosmic-ray species interact differently with grain material. This means that each CR element loses a different amount of energy ($E_{\rm lost}$, eV) upon impacting a grain. This energy also depends on the composition of grain, its density, size, and shape. The latter determines the exact cross-section and path length for CR particles striking the grain. Additionally, part of $E_{\rm lost}$ is carried away, when energetic electrons, released by the CR impact, escape from the finite-sized grain. The energy received by grains $E_{\rm grain}$ (eV) is transformed into heat and temporary raises grain temperature to $T_{CR}$. Each $T_{CR}$ occurs with a certain frequency $f_T$,~s$^{-1}$. The corresponding frequency for $E_{\rm grain}$ is $f_E$.

In the below sections, the above steps are described in detail. While each step has some uncertainties, we opted for the most plausible assortment of parameters to produce a single set of readily usable results. Along with the calculated $T_{\rm CR}$ spectra, the Appendices provide also $E_{\rm grain}$ spectra that allows to recalculate $T_{\rm CR}$ with a different approach on grain heat capacities.

\subsection{Cosmic-ray composition and energy spectrum}
\label{cr}
%
\begin{table}
\begin{center}
\caption{Adopted abundances of cosmic-ray constituents.}
\label{tab-ab}
\begin{tabular}{lc}
\tableline\tableline
X & [X]/[H] \\
\tableline
$^1$H & 1.00 \\
$^2$D & 1.60E-02 \\
$^3$He & 1.00E-02 \\
$^4$He & 8.36E-02 \\
Li & 9.65E-04 \\
Be & 1.16E-04 \\
B & 1.28E-03 \\
C & 5.19E-03 \\
N & 1.21E-03 \\
O & 5.01E-03 \\
F & 7.20E-05 \\
Ne & 7.06E-04 \\
Na & 1.34E-04 \\
Mg & 9.67E-04 \\
Al & 1.43E-04 \\
Si & 7.07E-04 \\
P & 1.85E-05 \\
S & 1.11E-04 \\
Cl & 1.76E-05 \\
Ar & 4.16E-05 \\
K & 2.94E-05 \\
Ca & 8.82E-05 \\
Sc & 1.84E-05 \\
Ti & 7.10E-05 \\
V & 3.23E-05 \\
Cr & 6.99E-05 \\
Mn & 4.34E-05 \\
Fe & 4.62E-04 \\
Co & 2.62E-06 \\
Ni & 1.97E-05 \\
\tableline
\end{tabular}
\end{center}
\end{table}
%
\begin{figure*}
 \vspace{-2cm}
  \hspace{-1cm}
  \includegraphics[width=18.0cm]{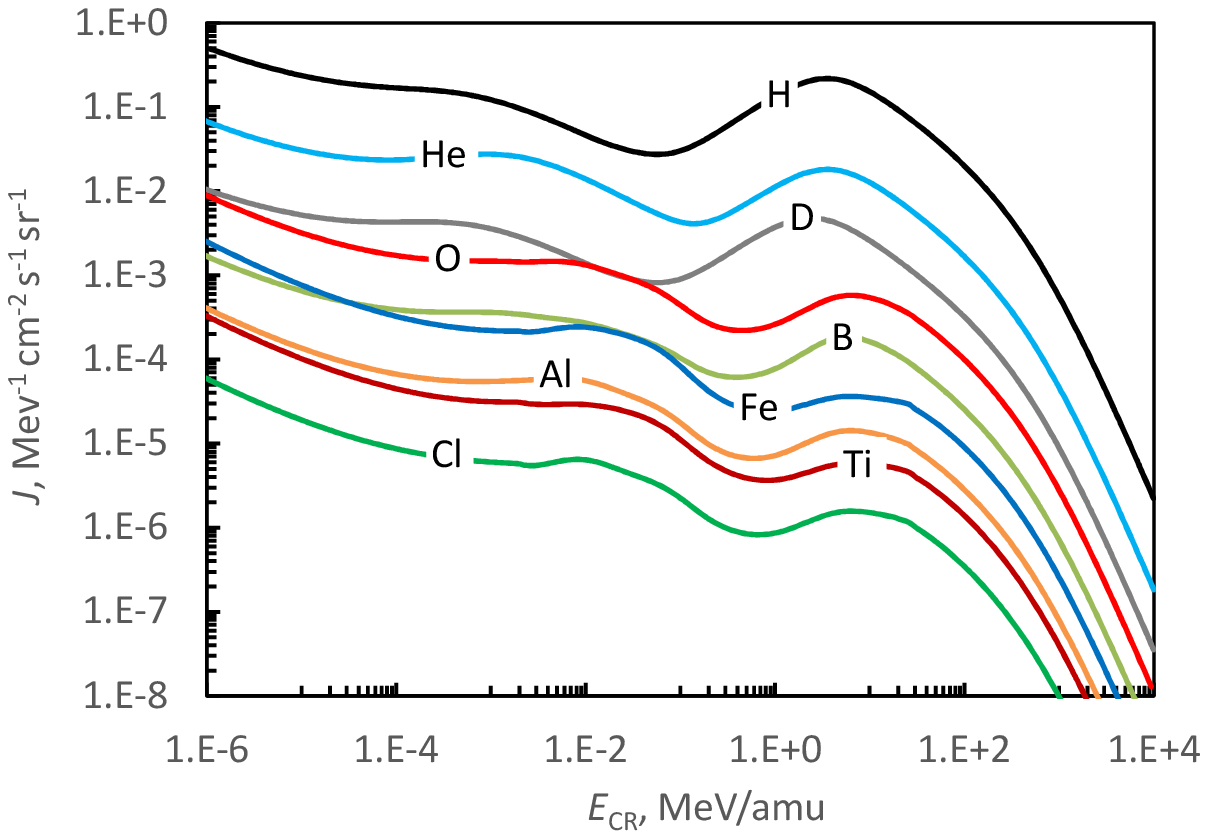}
 \vspace{-17.5cm}
 \caption{Calculated energy spectra for selected cosmic-ray components.}
 \label{att-en}
\end{figure*}
The CR energy spectrum was calculated generally following the work by \citet{Chabot16}. Cosmic-ray elements with atomic numbers from 1 to 28, i.e., from H to Ni, were considered. The relative abundances of the heavier elements starting with B were adopted from \citet{George09}. For estimating the relative abundance of hydrogen CRs, the Si/H ratio from \citet{Meyer98} was used. This approach results in that the important Fe/H abundance ratio falls approximately in the middle of the values used by other authors considering CR interactions with interstellar dust grains \citep{Leger85,Shen04,Roberts07,deBarros11d,Kalvans15a,Chabot16} and has an uncertainty of about a factor of two.

Regarding the lighter elements, we used a value of 8.36~\% for He/H abundance ratio from a dedicated study of light element CRs by \citet{Wang02}. These authors also indicate that CRs contain a considerable amount ($\approx1$~\% relative to H) of the isotopes D and $^3$He. The abundances of Li and Be were adopted from the Li/C and Be/C abundance ratios reported by \citet{deNolfo06}. Table~\ref{tab-ab} summarizes the adopted initial relative abundances of chemical elements in CRs.

It has been shown \citep{McCall03,Indriolo07,Indriolo09,Padovani09,Morlino15} that the flux of interstellar low-energy CRs is likely to be different and higher than earlier estimates \citep[e.g.,][]{Webber83} or the flux observed on Earth \citep{George09}. This can significantly affect grain heating because low-energy ions ($\approx1$~MeV/amu) are the most effective at depositing energy into the grains. Therefore, we adopted the ``High'' initial spectrum of CR protons recently devised by \citet{Ivlev15p}. This means that
   \begin{equation}
   \label{cr1}
J_{p,0} = 2.4\times10^{15}\frac{E^{-1.5}}{(E+5\times10^8~\rm{eV})^{1.9}}~\rm eV^{-1} cm^{-2} s^{-1} sr^{-1}.
   \end{equation}
Where $J_{p,0}$ is the initial spectrum of protons and $E$ is their energy per atomic mass unit. This spectrum was scaled by the abundances in Table~\ref{tab-ab} to obtain the initial spectra of the various other CR particles.

The CRs propagate through the interstellar cloud, interacting with its atoms. The resulting changes in the energy distribution of the CR particles were calculated according to
   \begin{equation}
   \label{cr2}
J_{k,N_H}=J_{k,0}\frac{L_k(E_{k,0})}{L_k(E_{k,N_H})},
   \end{equation}
where $J_{k,N_H}$ is the spectrum of species $k$ at a hydrogen column density $N_H$ (cm$^{-2}$) in the cloud and $E_k$ is measured in MeV/amu \citep{Padovani09}. Parameter $L_k$ is the CR energy loss function (eV per amount of matter), discussed below.

To calculate the CR spectrum, the value of $N_H$ has to be chosen. It is proportional to the interstellar visual extinction, $A_V$, measured in magnitudes. For the purposes of the present paper, two prerequisites have to be fulfilled: (1) that the chosen $A_V$ is characteristic for translucent molecular clouds and (2) that accumulation of ices is possible at the chosen $A_V$, so that our results are relevant to interstellar grain surface chemistry.

The range of extinctions, where these two conditions are met is fairly small, which allows us to employ a single $A_V$ value. Interstellar ice appears ar $A_V=3.2$~mag \citep{Whittet01}, which corresponds to an extinction of approximately 1.6~mag in the densest, central part of the cloud, where ice formation is taking place. Here we adopt a somewhat larger standard value of $A_V=2.0$~mag to account for the probability that at 1.6~mag the formation of ices may be just barely begun. Whole-grain heating at higher optical depths will be covered in a subsequent study.

To calculate the column density at $A_V=2.0$~mag, the value of the $N_H/A_V$ ratio has to be defined. The values suggested range from around $1.8\times10^{21}$ \citep{Reina73,Predehl95} to $2.2\times10^{21}$~cm$^{-2}$~mag$^{-1}$ \citep{Gorenstein75,Guver09}. Here we adopt a conventional middle value of $2.0\times10^{21}$~cm$^{-2}$~mag$^{-1}$.

Therefore, we assume that CR ions propagate through gas with hydrogen column density of $4.0\times10^{21}$~cm$^{-2}$. The loss function $L_k$ was calculated with the \textsc{srim}\footnote{http://www.srim.org/} package \citep{Ziegler85,Ziegler10} for a gas consisting of H, He ($\rm [He]/[H]=0.1$), and heavy elements C, N, O, Si, and Fe with relative abundances from \citet[parameter $F_\ast$ taken to be 0]{Jenkins09}. The calculated $L_k$ values are 1.2 times higher than those used by \citet{Chabot16}, mainly because of the inclusion of helium as a target gas component (note that for H we consider atoms, not molecules). Figure~\ref{att-en} shows the final spectrum of selected CR elements.

\subsection{Geometrical grain model}
\label{geom}
%
\begin{table}
\begin{center}
\caption{Assumed radius and ice thickness for the nine types of spherical interstellar grains considered in the study.}
\label{tab-sizes}
\begin{tabular}{clcc}
\tableline\tableline
 &  & Olivine core & Ice mantle \\
No. & Type & radius $a$, $\mu$m & thickness $b$, $\mu$m \\
\tableline
1 & Bare grain & 0.05 & - \\
2 & Thin ice & 0.05 & 0.005 \\
3 & Thick ice & 0.05 & 0.015 \\
4 & Bare grain & 0.1 & - \\
5 & Thin ice & 0.1 & 0.01 \\
6 & Thick ice & 0.1 & 0.03 \\
7 & Bare grain & 0.2 & - \\
8 & Thin ice & 0.2 & 0.02 \\
9 & Thick ice & 0.2 & 0.06 \\
\tableline
\end{tabular}
\end{center}
\end{table}
%
\begin{figure*}
\includegraphics[angle=270,width=5.0cm]{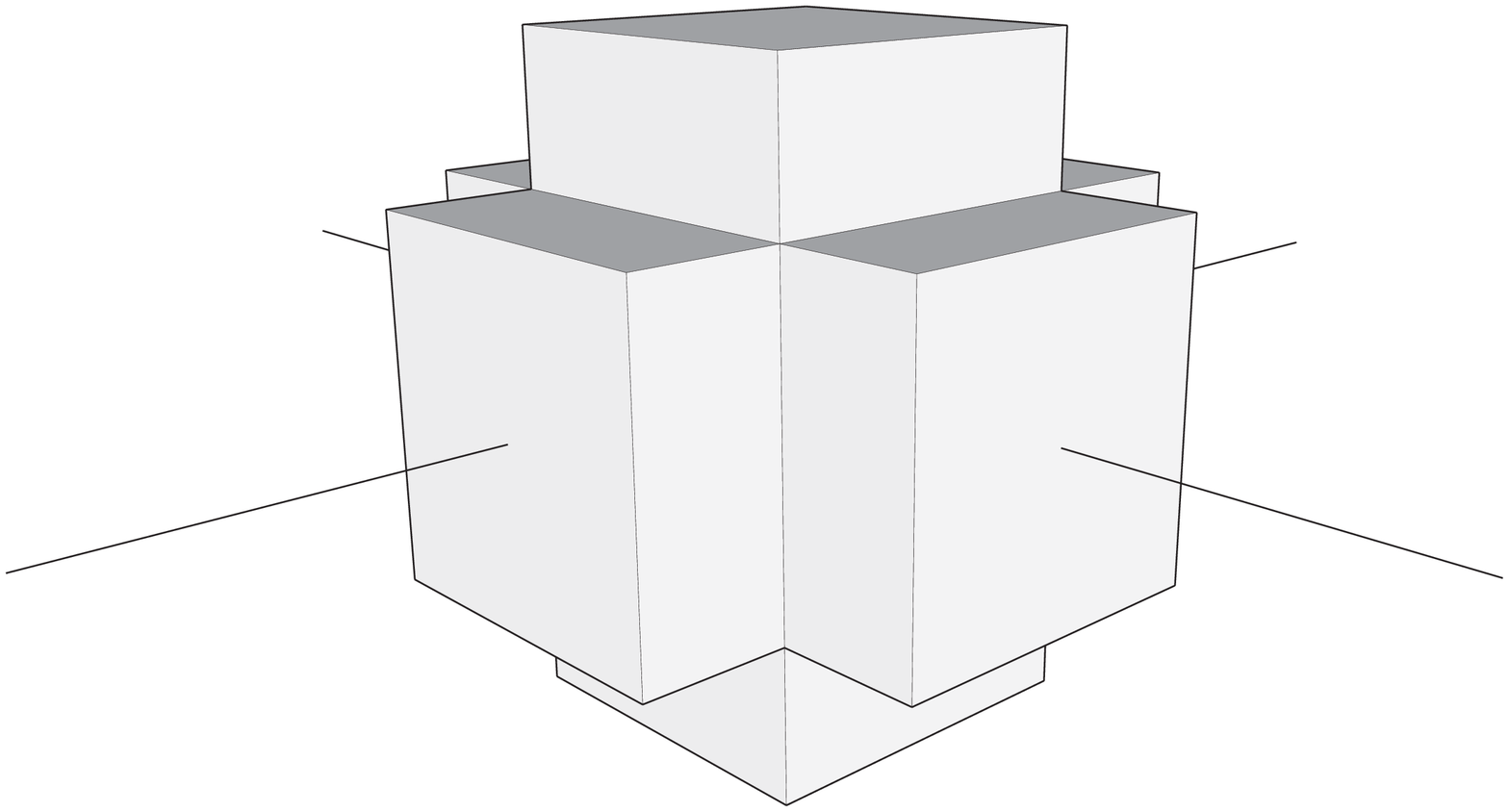}
\includegraphics[angle=270,width=5.0cm]{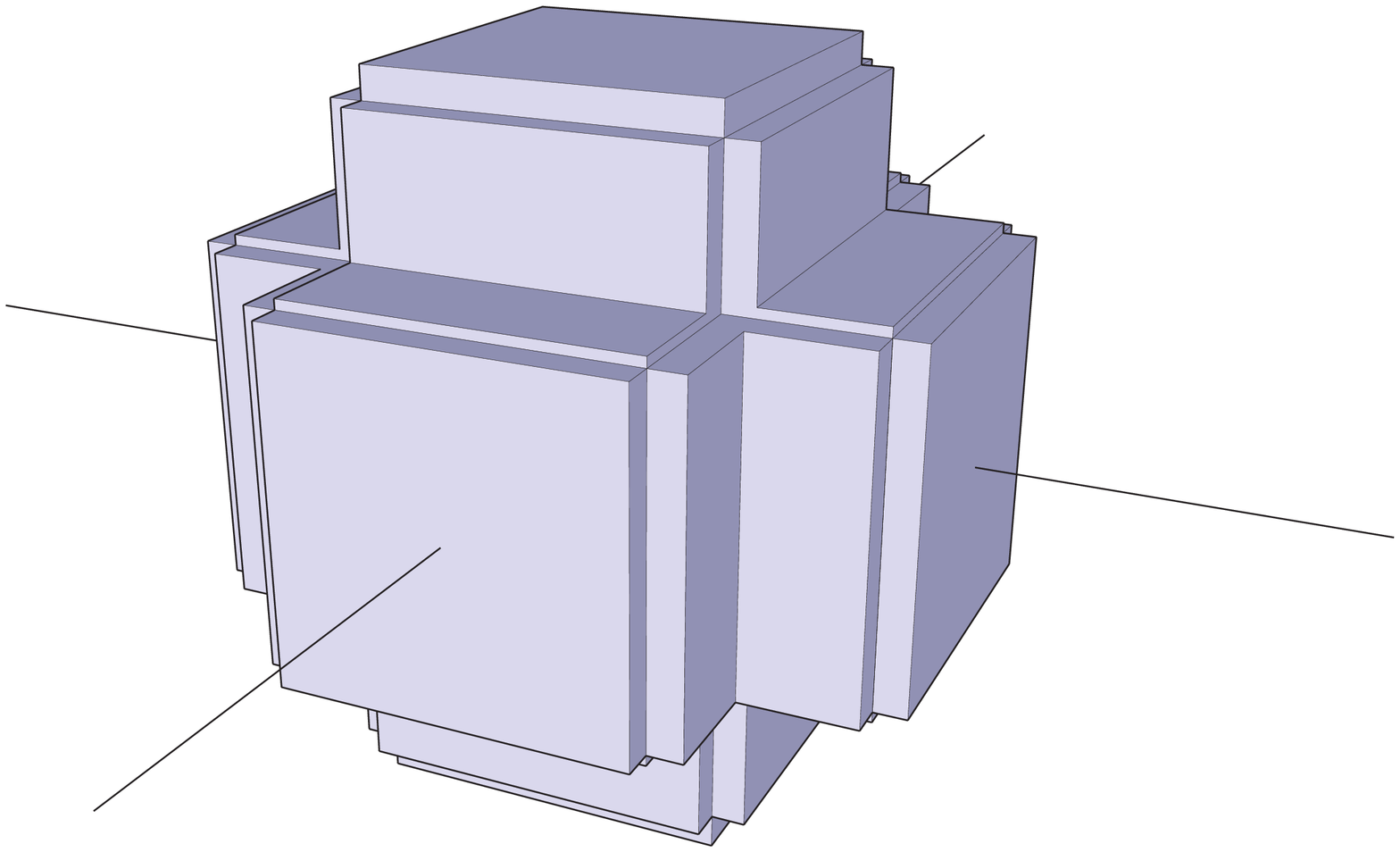}
\includegraphics[angle=270,width=5.0cm]{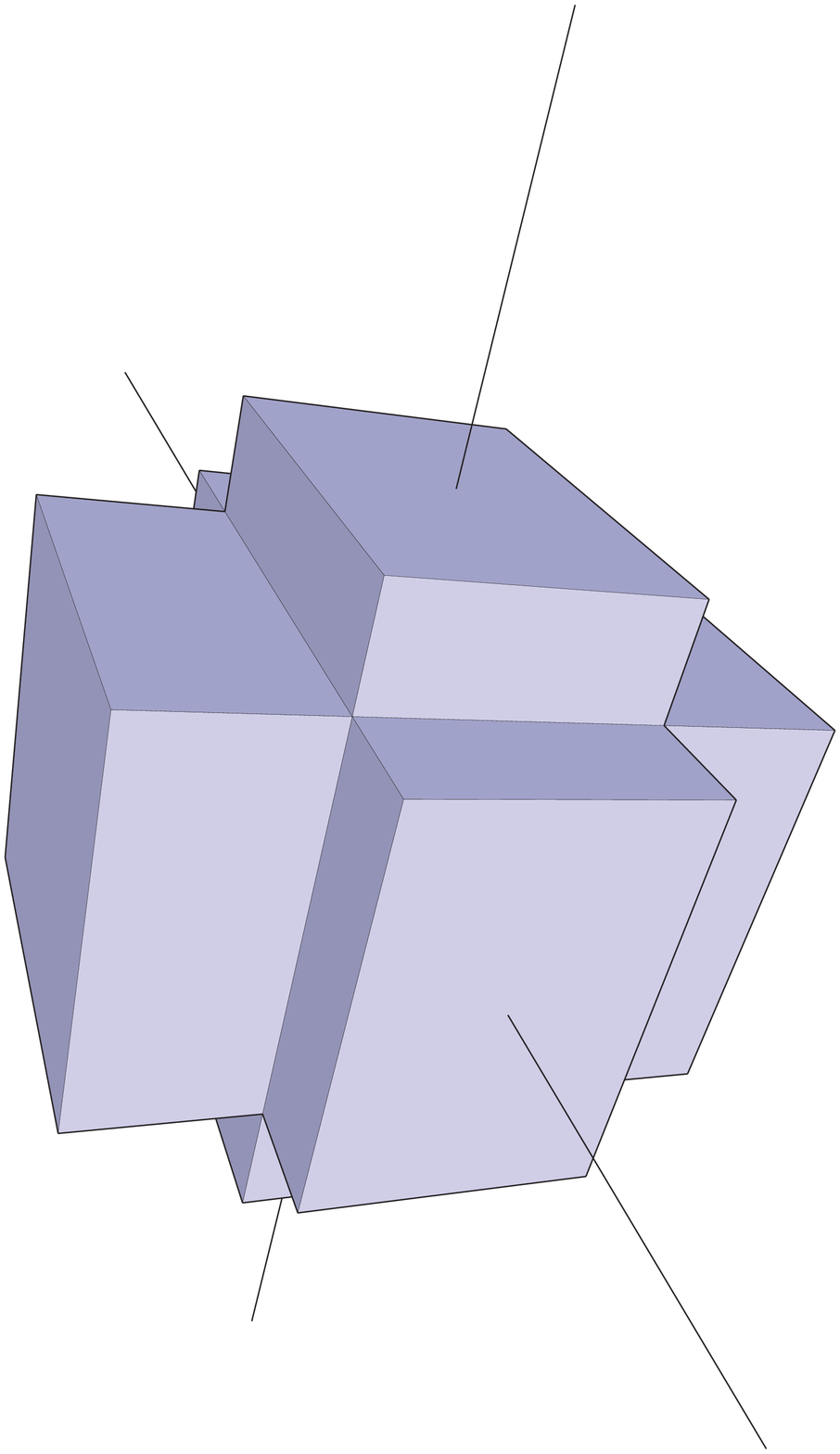}
\caption{Three-dimensional visualization of interstellar grain models, used for calculating cross-sections and path lengths for grain encounters with cosmic rays. From left to right: bare olivine grain, grain coated in thin ice mantle, and grain in thick ice mantle. The additional lines indicate the allowed directions of CRs, passing through the grain.}
\label{att-model}
\end{figure*}
%
\begin{figure*} 
 \vspace{-2cm}
 \hspace{-3cm}
\includegraphics{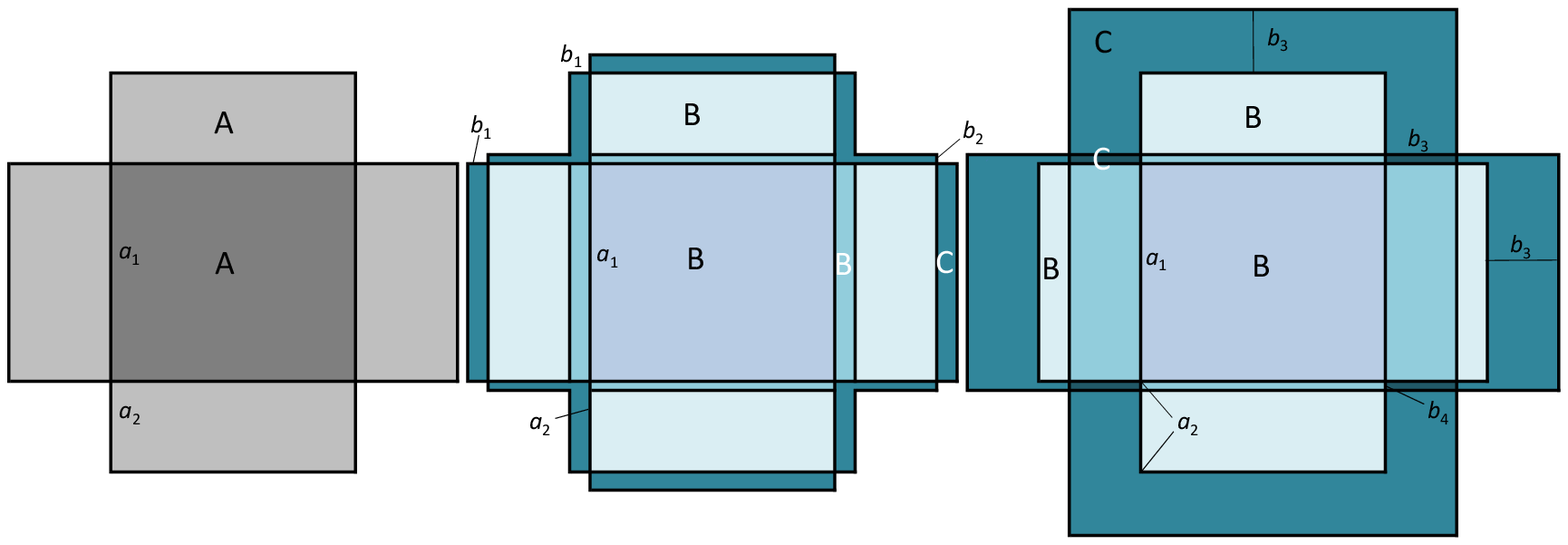}
 \vspace{-23cm}
\caption{Cross-sections for CR particle paths in bare grains, grains covered with a thin layer of ice, and those with a thick ice layer. Capital letters indicate the assumed possible CR paths through the grain: A--path through olivine; B--through olivine and ice; C--only through ice. The permitted paths are in the horizontal plane, either directly to or from the viewer, or perpendicularly to his line of sight. Lower case letters indicate the six parameters that characterize grain sizes--$a_1$ and $a_2$ for the bare olivine grains; $b_1$ and $b_2$ for the thin ice layer; $b_3$ and $b_4$ for the thick ice layer.}
\label{att-crossect}
\end{figure*}
Spherical grains with radius of $a=0.1\,\mu$m are typically adopted for astrochemical studies of molecular clouds. To enable the fitting of our data to a variety of environments, we consider grains with radius 0.05, 0.1, and 0.2~$\mu$m. The grain core material is olivine MgFeSiO$_4$. In dark cloud cores, the grains increase in size, probably because of accumulation of an ice layer \citep[e.g.,][]{Whittet01}. Therefore, for each grain size, we consider two additional cases: grains coated with a thin ice layer and grains in a thick ice mantle. Here, `thin ice' means ice with a uniform thickness of $b=0.1a$ and `thick ice'--with $b=0.3a$. For an average 0.1~$\mu$m grain, the thin and thick ice layers mean $\approx30$ and $\approx100$ ice molecule monolayers, corresponding to partial freeze-out in clouds and almost complete freeze-out in cloud cores, respectively. These ice thickness values are based on the results of our previous simulations of chemistry in contracting prestellar and starless cores \citep{Kalvans15a,Kalvans15c,Kalvans15b}. Table~\ref{tab-sizes} lists the basic parameters for the nine grain types considered.

We do not consider carbonaceous particles, such as PAHs, as grain constituents. These particles constitute $\approx10$~\% of grain mass \citep{Shen04}. Their light, carbon-dominated composition and relatively low density means that they are able to absorb little energy from passing CRs. Additionally, the heat capacity for carbonaceous material can be expected to be rather low, when compared to ices or silicates. It is also unclear to what extent such small carbonaceous particles are adsorbed onto dust grains in translucent clouds. Because the additional energy, absorbed by the carbonaceous materials from CRs and their additional heat capacity partially cancel each other out, we estimate that their contribution affects the $T_{\rm CR}$ spectrum by less than 10~\%.

Cosmic-ray particles pass through spherical interstellar grains with radius $a$ and ice layer thickness $b$. Each such CR particle transit has its path characterized by length $l$ through the grain and probability, proportional to the surface area d$S$, corresponding for each $l$. The energy deposited by the CR particle along its path through the grain is directly proportional to $l$. For grains with a core-mantle structure, $l$ can be further differentiated in path length through the olivine grain core $l_{\rm oli}\leq2a$ and through the icy mantle $l_{\rm ice}\geq2b$.

Previous authors \citep{Leger85,Shen04,Chabot16} assumed that the full cross-section ($\pi a^2$) of an interstellar grain corresponds to $l$ of one diameter, 2$a$. Practically, this means that the grains are assumed to be cylindrical and that all CR particles enter one of the ends of the cylinder, perpendicularly to its surface. Moreover, \citet{Shen04} considered icy grains and assumed that ice is evenly dispersed in the grain, without a core-mantle icy grain structure.

While adequate for the uncertainties regarding CRs and interstellar dust, the above approach can be improved. First, the volume of a cylindrical grain exceeds that of a sphere by a factor of 1.5. This overestimates the average CR path length and, therefore, the energy received by the grains because, in practice, the $E_{\rm grain}$ spectrum is affected by the existence of CR paths shorter than 2$a$. Second, an ice layer on the surface means that there are CR paths that do not touch the grain core. This is important because CRs deposit much less energy in the less-dense and light-element ices than in the olivine core. Thus, a path right through the center of the grain will generate much more heat than a path that merely touches the ice layer. These aspects have not yet been taken into account in studies considering CR-induced whole-grain heating.

We explored a number of possibilities on how to adequately reflect the interaction of spherical grains with CRs, while not producing an excessive amount of data. A convenient way is to assume a grain shape model that consists of a number of cuboids. To describe CR interaction with interstellar grains, a number of rules have to be obeyed for such a model: (1) the longest CR path must be 2$a$; (2) total area of the surface exposed to CRs must equal 2$\pi a^2$; (3) the total volume of the olivine core and the ice mantle must be close to that of a spherical grain. Rule (1) ensures that grain maximum temperatures are calculated properly, rule (2) means that the computed frequency of CR hits to the grains is correct, while rule (3) ensures that the average CR path length approximately matches that of a spherical grain (here, grain volume is the total grain cross-section multiplied by the average $l$). These three rules mean that the overall shape of the temperature spectrum for CR-heated grains and maximum temperatures are retained similar to the case of spherical grains. For grains with ice mantles, there are two additional rules: (4) $l_{\rm ice}$ must not be less than twice the thickness $b$ of the ice layer of a spherical grain and (5) CR path(s) through ice that do not touch the olivine core must be included.

Figure~\ref{att-model} displays the grain shapes employed for the calculation of cross-sections corresponding to specific CR path lengths. The edges of the olivine core model are characterized by parameters $a_1$ and $a_2$. Thin ice edges have two additional base edge lengths $b_1$ and $b_2$, while thick ice edge lengths are $b_3$ and $b_4$, as shown in Figure~\ref{att-crossect}. We assume that CRs may hit the grains only from certain directions, as indicated in the figures. There are two different CR paths for the bare grain model, five paths for the thin ice grain model, and six for grains with the thick ice layer.

The parameters $a_1$, $a_2$, $b_1$, $b_2$, $b_3$, and $b_4$ were chosen so that the rules (1-4) are fulfilled with a deviation of no more than $\pm6$~\%. Olivine and ice cross-sections are reproduced precisely by our model (margin of error $<0.2$~\%). Compared with the cylindrical grain models of earlier studies, the CR paths through the grain and the amount of energy deposited in the grain are now calculated more precisely.

\subsection{Transfer of energy from cosmic ray particles to grains}
\label{srim}
%
\begin{figure}
 \vspace{-3cm}
  \hspace{-1cm}
  \includegraphics[width=22.0cm]{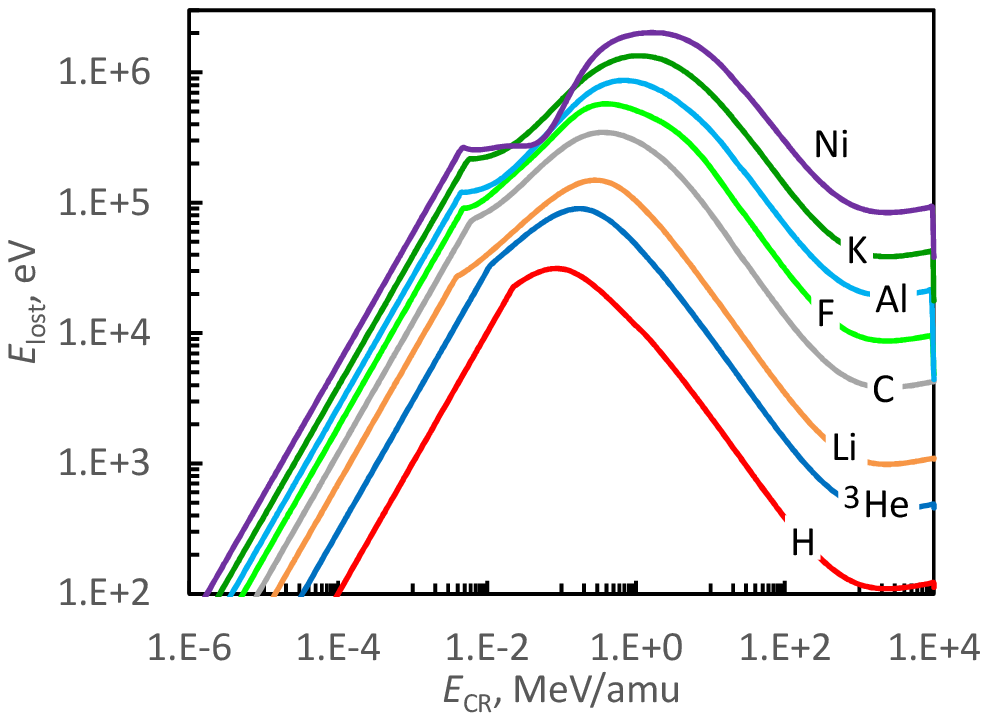}
 \vspace{-22.5cm}
 \caption{Energy lost by a selection of CR elemental particles when impacting the center part of a bare interstellar grain with radius $a=0.1\,\mu$m (i.e., passing through 0.2~$\mu$m of olivine). Data calculated with the \textsc{srim} program. The straight diagonal lines on the left-hand side indicate that the whole energy of the fast ion is lost upon impact.}
 \label{att-srim}
\end{figure}
%
\begin{figure}
 \vspace{-2cm}
  \hspace{-1cm}
  \includegraphics{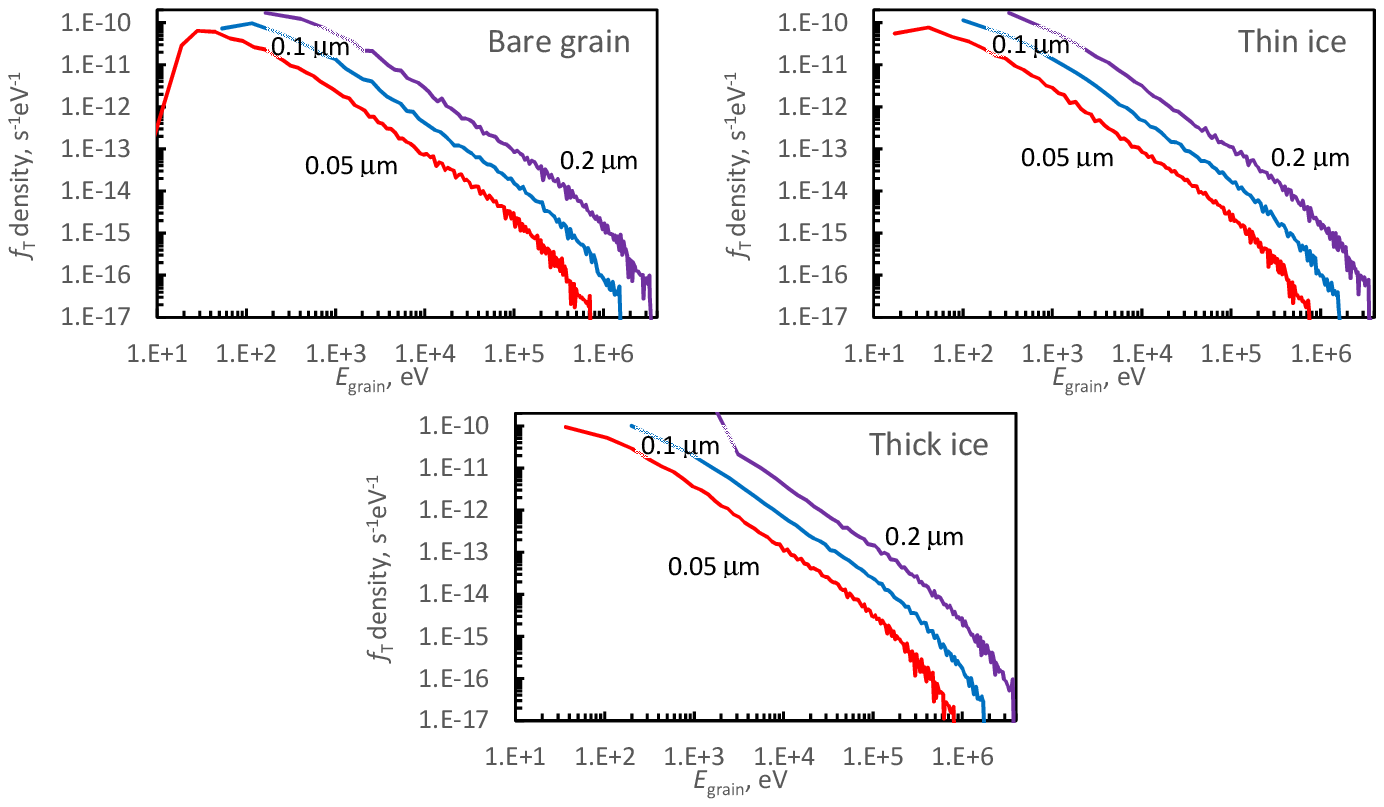}
 \vspace{-20cm}
 \caption{The spectrum of energy absorbed by interstellar grains from encounters with CRs. The frequency $f_E$ density per eV of CR collisions with grains is plotted versus the energy $E_{\rm grain}$ these encounters deliver to the grains. Data shown for bare grains, grains with a thin ice mantle and a thick ice mantle with olivine core sizes of 0.05, 0.1, and 0.2~$\mu$m.}
 \label{att-ev}
\end{figure}
As a CR nucleon transits an interstellar grain, it loses energy $E_{\rm lost}$ when interacting with electrons and atomic nuclei in the grain. The amount of $E_{\rm lost}$ depends on $l$, grain material, and density. We assumed that the grain core, with its geometry described above, is made of olivine MgFeSiO$_4$ (density 3.32~g~cm$^{-3}$). The ice has an assumed density of 1.0~g~cm$^{-3}$ and consists of H$_2$O, CO, CO$_2$, and CH$_3$OH with abundance ratios 100:31:38:4 \citep{Oberg11}. For a CR particle that travels a distance through olivine and ice, $E_{\rm lost}$ was calculated with the \textsc{srim} package. Electronic and nuclear interactions both were included. Figure~\ref{att-srim} presents examples for the amount of energy lost by selected CR elements when passing through the center part of a 0.1~$\mu$m bare olivine grain.

As a CR particle zips through the grain, it releases electrons from grain material and a cylinder along the fast ion's path becomes heated and ionized. This occurs in a picosecond timescale \citep{Iza06}. Some of the electrons released by the impact escape the grain. The proportion of energy carried away with the escaping electrons was calculated according to the data from \citet{Leger85}. The remaining energy is converted into heat. Within $\approx10^{-10}$~s the heat diffuses to the whole grain \citep{Leger85}. For the purpose of the the present study, we assume that all of the energy deposited in the grain ($E_{\rm grain}$) instantly transforms into heat, before any substantial cooling occurs.

Figure~\ref{att-ev} shows the frequency $f_E$ density (s$^{-1}$eV$^{-1}$) spectrum of energy $E_{\rm grain}$ received by grains of different radii and ice thickness. These data, presented in tabulated form in Appendix~\ref{app-a}, were calculated as explained in the above sections and were used for the calculation of the $T_{\rm CR}$ spectra in Section~\ref{wgh}. The `energy frequency' $f_E$ corresponds to a certain `temperature frequency' $f_T$, with grain heat capacity as the conversion factor between $E_{\rm grain}$ and $T_{\rm CR}$.

\subsection{Calculation of grain temperatures}
\label{wgh}
%
\begin{figure}
 \vspace{-2cm}
  \hspace{3cm}
  \includegraphics{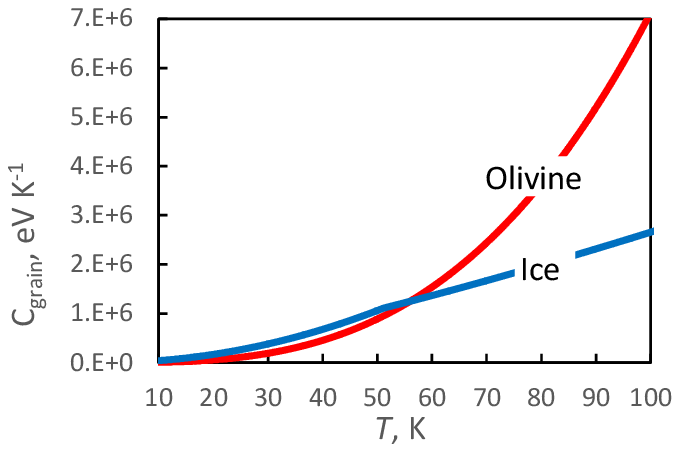}
 \vspace{-24cm}
 \caption{Comparison of heat capacities for olivine and ice spheres with a radius of 0.1~$\mu$m.}
 \label{att-cgrain}
\end{figure}
%
\begin{figure*}
 \vspace{-3cm}
  \hspace{-2cm}
  \includegraphics{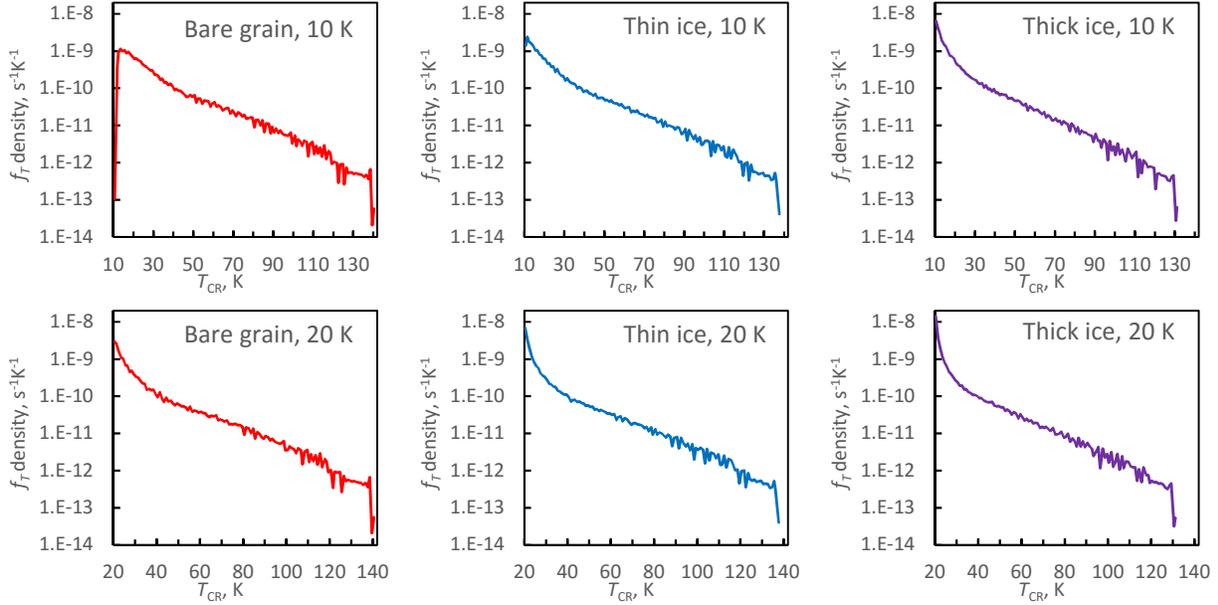}
 \vspace{-21cm}
 \caption{The temperature $T_{\rm CR}$ spectrum of grains with $a=0.05\,\mu$m, heated by CR particles. The frequency $f_T$ density per K of CR encounters with grains is plotted versus the temperature $T_{\rm CR}$ achieved by the grains due to impacts by CR particles.}
 \label{att-tcr0.05}
\end{figure*}
%
\begin{figure*}
 \vspace{-3cm}
  \hspace{-2cm}
  \includegraphics{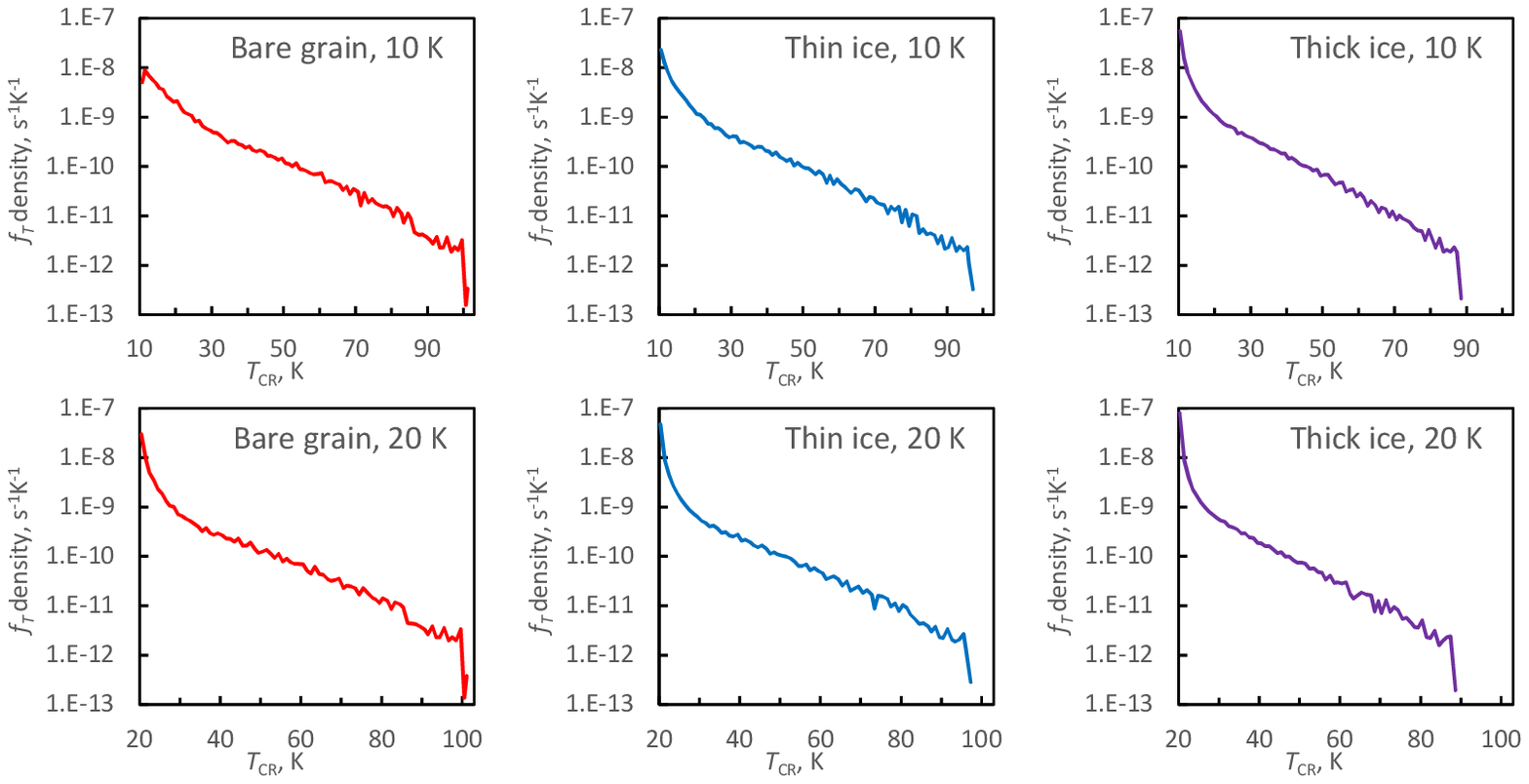}
 \vspace{-21cm}
 \caption{The temperature $T_{\rm CR}$ spectrum of grains with $a=0.1\,\mu$m, heated by CR particles.}
 \label{att-tcr0.1}
\end{figure*}
%
\begin{figure*}
 \vspace{-2cm}
  \hspace{-2cm}
  \includegraphics{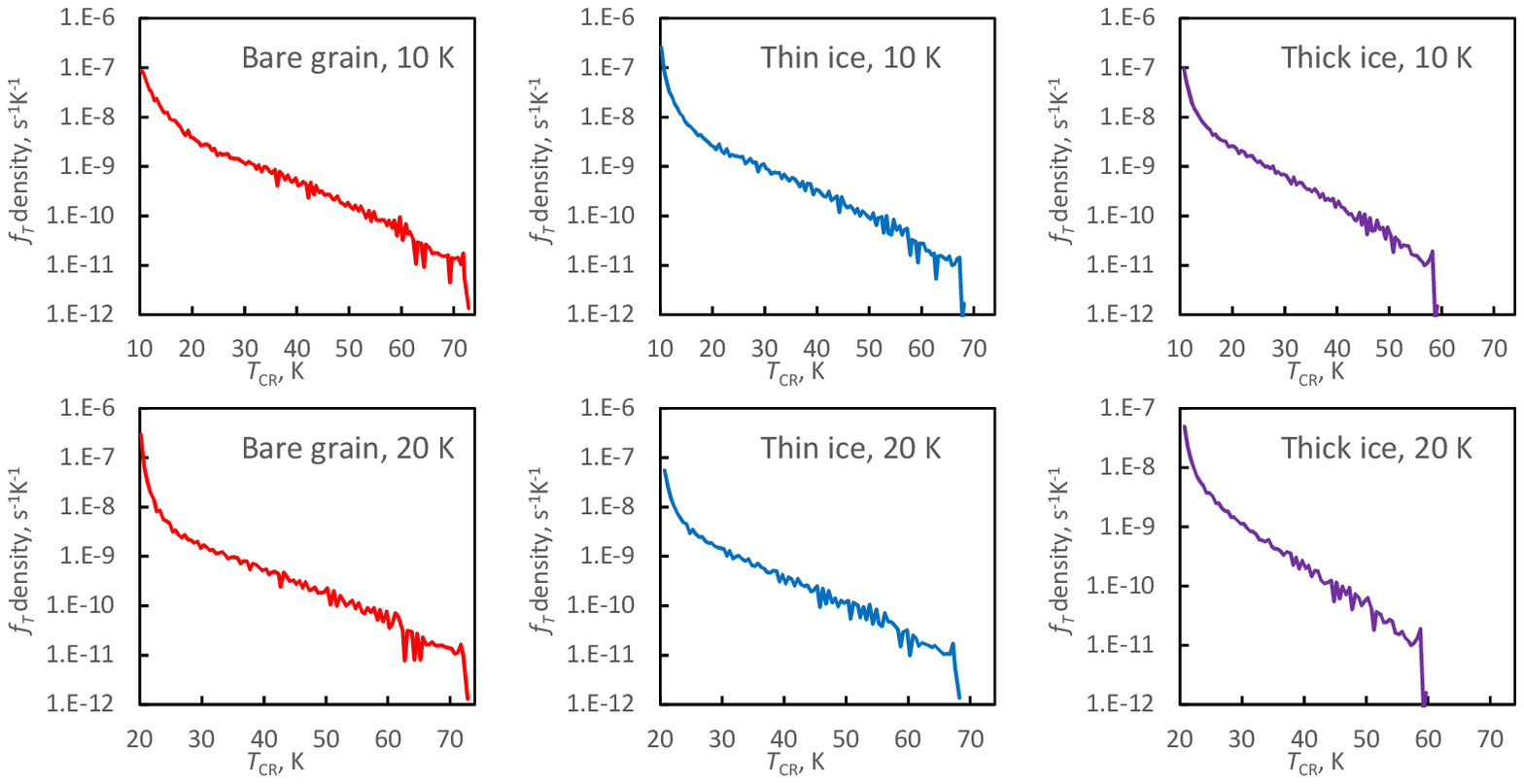}
 \vspace{-21cm}
 \caption{The temperature $T_{\rm CR}$ spectrum of grains with $a=0.2\,\mu$m, heated by CR particles.}
 \label{att-tcr0.2}
\end{figure*}
For a grain that has received a certain energy amount $E_{\rm grain}$, the reached temperature $T_{\rm CR}$ depends on heat capacity $C_{\rm grain}$ (eV~K$^{-1}$) and initial (equilibrium) grain temperature $T_0$. For the later, 10~K can be regarded as a typical value in molecular clouds, which also has often been employed in astrochemical simulations. Therefore, we take 10~K as a valid `standard' value for $T_0$. Heat capacities at temperatures below 10~K are very low. Because of this, 10~K is representative also for dark and dense cores, where $T_0$ can be as low as 5~K. For more diffuse gas and cloud cores affected by some kind of energy influx, we adopt a second value for $T_0=20$~K. Thus, nine grain types (Table~\ref{tab-sizes}) and two $T_0$ values of 10 and 20~K bring the total number of data sets to 18.

Whole-grain heating for grains with $T_0>20$~K was not included because such heating has a limited chemical effect. For example, a process with an activation energy $E_A$ barrier of 1000~K can be enabled by whole-grain heating of 10~K grains. The same process on 20~K grains occurs with an already appreciable rate without additional energy influx. From our presented data, we estimate that processes with $1200<E_A<1500$~K could be enabled by CR-induced heating of 20~K grains. For grains with even higher $T_0$, such $E_A$ range is even narrower. Additionally, warmer grains, when heated, do not reach significantly higher $T_{\rm CR}$ values than cold grains, because the higher the temperature, the higher is grain heat capacity (Figure~\ref{att-cgrain}).

We calculate the heat capacity for the olivine grains $C_{\rm grain}$ according to Eq.~(18) of \citet{Cuppen06}. Following \citet{Leger85}, who indicate that the volumnic heat capacity of amorphous water ice at lower temperatures is slightly higher than that of silicates, we calculate the heat capacity of the ice layer according to the latter authors' approach. Here, we use the geometry of spherical grains coated with icy mantles of uniform thickness, not the grain model outlined in Section~\ref{geom}. Figure~\ref{att-cgrain} shows a comparison of heat capacities for fictional olivine and ice particles.

With the acquisition of an ice mantle, the grains get larger, which results in more frequent CR hits. These additional hits touch only the ice layer and do not deposit high amounts of energy in the grain. However, the additional heat capacity of the icy mantle is considerable even for grains with a thin ice layer. As a consequence, we find that the occurrence frequency $f_T$ for a certain value of $T_{\rm CR}$ is always lowered with the addition of the ice layer to a grain.

The full calculated whole-grain heating temperature spectra is listed in the Appendices. Appendix~\ref{app-a} lists also the energies $E_{\rm grain}$ that allow to recalculate the $T_{\rm CR}$ spectrum for various grain equilibrium temperatures and different approaches on heat capacities.

In the calculations, temperatures were rounded to a precision of 0.01~K. The $T_{\rm CR}$ spectrum is best represented by approximately hundred data points because we calculate $E_{\rm grain}$ for a finite number of ion energies, representing the CR energy spectra. The $\approx$hundred entries correspond to 1~K temperature intervals for 0.05 and 0.1~$\mu$m grains and 0.5~K intervals for grains with $a=0.2\,\mu$m. We found that this approach resolves the $T_{\rm CR}$ spectrum with its characteristic major features in sufficient detail (Figures~\ref{att-tcr0.05}--\ref{att-tcr0.2}).

\subsection{Uncertainties and application limits}
\label{uncer}

Uncertainties in the calculated temperatures $T_{\rm CR}$ arise from approximations in grain interaction with CR particles (Section~\ref{srim}) and grain material properties (\ref{wgh}). We estimate that their margins of error do not exceed factor of 1.5. Each of these aspects was treated in a more detailed manner than in previous studies considering whole-grain heating \citep{Leger85,Hasegawa93,Roberts07}. The latter statement is true also for grain geometry (\ref{geom}), where errors are relatively negligible.

Uncertainties in $f_T$ arise from the calculated CR spectra (Section~\ref{cr}), which largely depend on the loss function and less so on the adopted initial low-energy spectrum \citep{Padovani09,Morlino15}. The margin of error for the final spectra probably lies within a factor of few. Additionally, while a cloud with $A_V=2$~mag was considered, the calculated CR fluxes change within a factor of two in $A_V$ range of 1...5~mag. The above means that, for the case of translucent clouds, the calculated $T_{\rm CR}$ spectra have a maximum uncertainty of about an order of magnitude, comparable to that of CR-induced ionization rate in diffuse clouds.

\section{Results and discussion}
\label{disco}
%
\begin{table*}
\begin{center}
\caption{Frequency of CR encounters with grains ($T_0=10$~K) that lift the grain temperature above a certain minimum $T_{\rm CR}$ threshold. The summed-over frequency $\Sigma f_T$ is in units of s$^{-1}$ and the average time $t_{\rm CR}$ between CR hits to the grain is in years.}
\label{tab-yr10}
\resizebox{\linewidth}{!}{
\begin{tabular}{llccccccccc}
\tableline\tableline
 &  & 0.05 $\mu$m & 0.05 $\mu$m & 0.05 $\mu$m & 0.1 $\mu$m & 0.1 $\mu$m & 0.1 $\mu$m & 0.2 $\mu$m & 0.2 $\mu$m & 0.2 $\mu$m \\
$T_{\rm CR}$, K &  & bare grain & thin ice & thick ice & bare grain & thin ice & thick ice & bare grain & thin ice & thick ice \\
\tableline
$>20$ & $\Sigma f_T$ & 8.14E-09 & 6.51E-09 & 5.54E-09 & 1.76E-08 & 1.29E-08 & 1.07E-08 & 3.39E-08 & 2.46E-08 & 1.85E-08 \\
 & $t_{\rm CR}$ & 3.89E+00 & 4.86E+00 & 5.72E+00 & 1.80E+00 & 2.46E+00 & 2.97E+00 & 9.35E-01 & 1.29E+00 & 1.72E+00 \\
$>27$ & $\Sigma f_T$ & 4.54E-09 & 3.61E-09 & 3.16E-09 & 8.81E-09 & 6.96E-09 & 5.68E-09 & 1.66E-08 & 1.17E-08 & 7.11E-09 \\
 & $t_{\rm CR}$ & 6.98E+00 & 8.77E+00 & 1.00E+01 & 3.60E+00 & 4.55E+00 & 5.58E+00 & 1.91E+00 & 2.71E+00 & 4.46E+00 \\
$>30$ & $\Sigma f_T$ & 3.59E-09 & 2.94E-09 & 2.59E-09 & 7.01E-09 & 5.61E-09 & 4.36E-09 & 1.25E-08 & 8.28E-09 & 4.69E-09 \\
 & $t_{\rm CR}$ & 8.82E+00 & 1.08E+01 & 1.22E+01 & 4.52E+00 & 5.65E+00 & 7.26E+00 & 2.54E+00 & 3.83E+00 & 6.76E+00 \\
$>40$ & $\Sigma f_T$ & 1.91E-09 & 1.66E-09 & 1.43E-09 & 3.51E-09 & 2.66E-09 & 1.78E-09 & 4.28E-09 & 2.57E-09 & 1.08E-09 \\
 & $t_{\rm CR}$ & 1.66E+01 & 1.91E+01 & 2.22E+01 & 9.02E+00 & 1.19E+01 & 1.78E+01 & 7.41E+00 & 1.23E+01 & 2.93E+01 \\
$>50$ & $\Sigma f_T$ & 1.17E-09 & 9.94E-10 & 7.99E-10 & 1.67E-09 & 1.20E-09 & 7.05E-10 & 1.28E-09 & 6.91E-10 & 1.76E-10 \\
 & $t_{\rm CR}$ & 2.72E+01 & 3.19E+01 & 3.97E+01 & 1.90E+01 & 2.63E+01 & 4.50E+01 & 2.47E+01 & 4.59E+01 & 1.80E+02 \\
$>60$ & $\Sigma f_T$ & 7.09E-10 & 5.85E-10 & 4.39E-10 & 7.50E-10 & 5.07E-10 & 2.53E-10 & 2.74E-10 & 1.15E-10 & \nodata \\
 & $t_{\rm CR}$ & 4.47E+01 & 5.41E+01 & 7.22E+01 & 4.23E+01 & 6.25E+01 & 1.25E+02 & 1.16E+02 & 2.75E+02 & \nodata \\
$>70$ & $\Sigma f_T$ & 4.11E-10 & 3.40E-10 & 2.42E-10 & 3.04E-10 & 2.03E-10 & 8.66E-11 & 3.14E-11 & \nodata & \nodata \\
 & $t_{\rm CR}$ & 7.70E+01 & 9.33E+01 & 1.31E+02 & 1.04E+02 & 1.56E+02 & 3.66E+02 & 1.01E+03 & \nodata & \nodata \\
$>80$ & $\Sigma f_T$ & 2.30E-10 & 1.90E-10 & 1.27E-10 & 1.07E-10 & 6.81E-11 & 1.93E-11 & \nodata & \nodata & \nodata \\
 & $t_{\rm CR}$ & 1.38E+02 & 1.67E+02 & 2.50E+02 & 2.96E+02 & 4.65E+02 & 1.64E+03 & \nodata & \nodata & \nodata \\
$>90$ & $\Sigma f_T$ & 1.25E-10 & 1.02E-10 & 6.54E-11 & 2.78E-11 & 1.58E-11 & \nodata & \nodata & \nodata & \nodata \\
 & $t_{\rm CR}$ & 2.54E+02 & 3.11E+02 & 4.85E+02 & 1.14E+03 & 2.01E+03 & \nodata & \nodata & \nodata & \nodata \\
$>100$ & $\Sigma f_T$ & 6.46E-11 & 5.29E-11 & 3.18E-11 & 2.30E-13 & \nodata & \nodata & \nodata & \nodata & \nodata \\
 & $t_{\rm CR}$ & 4.90E+02 & 5.99E+02 & 9.97E+02 & 1.38E+05 & \nodata & \nodata & \nodata & \nodata & \nodata \\
$>110$ & $\Sigma f_T$ & 3.01E-11 & 2.37E-11 & 1.32E-11 & \nodata & \nodata & \nodata & \nodata & \nodata & \nodata \\
 & $t_{\rm CR}$ & 1.05E+03 & 1.34E+03 & 2.40E+03 & \nodata & \nodata & \nodata & \nodata & \nodata & \nodata \\
$>120$ & $\Sigma f_T$ & 1.11E-11 & 9.22E-12 & 4.05E-12 & \nodata & \nodata & \nodata & \nodata & \nodata & \nodata \\
 & $t_{\rm CR}$ & 2.86E+03 & 3.44E+03 & 7.82E+03 & \nodata & \nodata & \nodata & \nodata & \nodata & \nodata \\
$>130$ & $\Sigma f_T$ & 4.38E-12 & 2.88E-12 & 4.72E-14 & \nodata & \nodata & \nodata & \nodata & \nodata & \nodata \\
 & $t_{\rm CR}$ & 7.23E+03 & 1.10E+04 & 6.71E+05 & \nodata & \nodata & \nodata & \nodata & \nodata & \nodata \\
$>140$ & $\Sigma f_T$ & 3.76E-14 & \nodata & \nodata & \nodata & \nodata & \nodata & \nodata & \nodata & \nodata \\
 & $t_{\rm CR}$ & 8.44E+05 & \nodata & \nodata & \nodata & \nodata & \nodata & \nodata & \nodata & \nodata \\
\tableline
\end{tabular}
}
\end{center}
\end{table*}
%
\begin{table*}
\begin{center}
\footnotesize
\caption{Frequency of CR encounters with grains ($T_0=20$~K) that lift the grain temperature above a certain $T_{\rm CR}$ threshold. $\Sigma f_T$ is in units of s$^{-1}$ and $t_{\rm CR}$ is in years.}
\label{tab-yr20}
\resizebox{\linewidth}{!}{
\begin{tabular}{llccccccccc}
\tableline\tableline
 &  & 0.05 $\mu$m & 0.05 $\mu$m & 0.05 $\mu$m & 0.1 $\mu$m & 0.1 $\mu$m & 0.1 $\mu$m & 0.2 $\mu$m & 0.2 $\mu$m & 0.2 $\mu$m \\
$T_{\rm CR}$, K &  & bare grain & thin ice & thick ice & bare grain & thin ice & thick ice & bare grain & thin ice & thick ice \\
\tableline
$>30$ & $\Sigma f_T$ & 4.05E-09 & 3.39E-09 & 3.04E-09 & 7.85E-09 & 6.46E-09 & 5.40E-09 & 1.44E-08 & 1.04E-08 & 6.61E-09 \\
 & $t_{\rm CR}$ & 7.83E+00 & 9.34E+00 & 1.04E+01 & 4.03E+00 & 4.91E+00 & 5.87E+00 & 2.20E+00 & 3.04E+00 & 4.80E+00 \\
$>40$ & $\Sigma f_T$ & 1.98E-09 & 1.72E-09 & 1.53E-09 & 3.66E-09 & 2.84E-09 & 1.99E-09 & 4.59E-09 & 2.87E-09 & 1.30E-09 \\
 & $t_{\rm CR}$ & 1.60E+01 & 1.84E+01 & 2.07E+01 & 8.66E+00 & 1.11E+01 & 1.59E+01 & 6.90E+00 & 1.10E+01 & 2.44E+01 \\
$>50$ & $\Sigma f_T$ & 1.18E-09 & 1.02E-09 & 8.33E-10 & 1.73E-09 & 1.26E-09 & 7.54E-10 & 1.35E-09 & 7.39E-10 & 2.10E-10 \\
 & $t_{\rm CR}$ & 2.68E+01 & 3.12E+01 & 3.81E+01 & 1.83E+01 & 2.52E+01 & 4.20E+01 & 2.35E+01 & 4.29E+01 & 1.51E+02 \\
$>60$ & $\Sigma f_T$ & 7.16E-10 & 5.95E-10 & 4.47E-10 & 7.57E-10 & 5.24E-10 & 2.66E-10 & 2.84E-10 & 1.17E-10 & \nodata \\
 & $t_{\rm CR}$ & 4.43E+01 & 5.33E+01 & 7.09E+01 & 4.18E+01 & 6.04E+01 & 1.19E+02 & 1.12E+02 & 2.71E+02 & \nodata \\
$>70$ & $\Sigma f_T$ & 4.15E-10 & 3.43E-10 & 2.45E-10 & 3.06E-10 & 2.08E-10 & 8.94E-11 & 3.17E-11 & \nodata & \nodata \\
 & $t_{\rm CR}$ & 7.63E+01 & 9.23E+01 & 1.29E+02 & 1.03E+02 & 1.52E+02 & 3.55E+02 & 9.99E+02 & \nodata & \nodata \\
$>80$ & $\Sigma f_T$ & 2.30E-10 & 1.91E-10 & 1.30E-10 & 1.12E-10 & 6.90E-11 & 2.12E-11 & \nodata & \nodata & \nodata \\
 & $t_{\rm CR}$ & 1.38E+02 & 1.66E+02 & 2.44E+02 & 2.82E+02 & 4.59E+02 & 1.50E+03 & \nodata & \nodata & \nodata \\
$>90$ & $\Sigma f_T$ & 1.27E-10 & 1.02E-10 & 6.57E-11 & 2.78E-11 & 1.58E-11 & \nodata & \nodata & \nodata & \nodata \\
 & $t_{\rm CR}$ & 2.50E+02 & 3.09E+02 & 4.82E+02 & 1.14E+03 & 2.01E+03 & \nodata & \nodata & \nodata & \nodata \\
$>100$ & $\Sigma f_T$ & 6.46E-11 & 5.31E-11 & 3.19E-11 & 2.35E-13 & \nodata & \nodata & \nodata & \nodata & \nodata \\
 & $t_{\rm CR}$ & 4.90E+02 & 5.97E+02 & 9.94E+02 & 1.35E+05 & \nodata & \nodata & \nodata & \nodata & \nodata \\
$>110$ & $\Sigma f_T$ & 3.01E-11 & 2.38E-11 & 1.34E-11 & \nodata & \nodata & \nodata & \nodata & \nodata & \nodata \\
 & $t_{\rm CR}$ & 1.05E+03 & 1.33E+03 & 2.36E+03 & \nodata & \nodata & \nodata & \nodata & \nodata & \nodata \\
$>120$ & $\Sigma f_T$ & 1.11E-11 & 9.23E-12 & 4.39E-12 & \nodata & \nodata & \nodata & \nodata & \nodata & \nodata \\
 & $t_{\rm CR}$ & 2.86E+03 & 3.43E+03 & 7.22E+03 & \nodata & \nodata & \nodata & \nodata & \nodata & \nodata \\
$>130$ & $\Sigma f_T$ & 4.39E-12 & 2.88E-12 & 5.19E-14 & \nodata & \nodata & \nodata & \nodata & \nodata & \nodata \\
 & $t_{\rm CR}$ & 7.23E+03 & 1.10E+04 & 6.11E+05 & \nodata & \nodata & \nodata & \nodata & \nodata & \nodata \\
$>140$ & $\Sigma f_T$ & 3.76E-14 & \nodata & \nodata & \nodata & \nodata & \nodata & \nodata & \nodata & \nodata \\
 & $t_{\rm CR}$ & 8.44E+05 & \nodata & \nodata & \nodata & \nodata & \nodata & \nodata & \nodata & \nodata \\
\tableline
\end{tabular}
}
\end{center}
\end{table*}
Appendix~\ref{app-a} presents the temperature $T_{\rm CR}$ spectra for CR-induced whole-grain heating of $T_0=10$~K interstellar grains with properties described above (Sections \ref{geom} and \ref{srim}). Tables of Appendix~\ref{app-b} list the $T_{\rm CR}$ spectra for 20~K grains. We remind that the nine grain types include grains with olivine core radius $a$ 0.05, 0.1, and 0.2~$\mu$m; bare grains, and grains with thin and thick icy mantles. The data of temperature spectra are shown graphically in Figures \ref{att-tcr0.05}--\ref{att-tcr0.2}.

To ensure a safe reproduction of our calculated temperature spectra from data in the Appendices, for each entry, we state the $T_{\rm CR}$ interval, its width $\Delta T_{\rm CR}$, and the corresponding weighed average $T_{\rm CR}$ value for each interval. The corresponding $T_{\rm CR}$ occurrence frequency $f_T$ is the sum of impact frequencies for CR nuclei H through Ni that deliver the energy $E_{\rm grain}$ required to lift grain temperature to a $T_{\rm CR}$ value, which falls in the specified temperature interval. For example, a grain with radius $a=0.1\, \mu$m, ice mantle thickness $b=0.03\,\mu$m, and $T_0=10$~K can reach temperatures in the interval 32.01--33.00~K due of hits by He nuclei with $E_{\rm CR}=1.7$~MeV/ion that pass right through the center of the grain, due to hits by 9~GeV/ion Ni nuclei that touch only the ice layer, and due to hits by all CR components in between.

Appendix~\ref{app-a} includes also data on the received energy $E_{\rm grain}$ spectrum for all the grain types in consideration, as shown in Figure~\ref{att-ev}. This energy is received by the grain from a CR hit and has a corresponding grain temperature $T_{\rm CR}$. We specify the weighed average $E_{\rm grain}$, its corresponding energy interval (pay attention to footnote b below Table~\ref{tab-A1}), and interval width $\Delta E_{\rm grain}$. In the specified $T_{\rm CR}$ and $E_{\rm grain}$ intervals of each entry, $f_T$ equals $f_E$ (see also Section~\ref{srim}). The conversion factors between $E_{\rm grain}$ and $T_{\rm CR}$ is the Boltzmann constant and the total grain heat capacity $C_{T_0-T_{\rm CR}}$ (eV) in the relevant temperature interval from $T_0$ to $T_{\rm CR}$:
   \begin{equation}
   \label{phys1}
C_{T_0-T_{\rm CR}}=\int^{T_0}_{T_{\rm CR}}C_{\rm{grain}}{\rm{d}}T.
   \end{equation}
Because there can be more than one variant on how to calculate grain heat capacity $C_{\rm grain}$, the $E_{\rm grain}$ frequency spectrum allows to recalculate the $T_{\rm CR}$ spectrum with a different approach on $C_{\rm grain}$. We do not provide the $E_{\rm grain}$ data in Appendix~\ref{app-b} because this quantity is independent of temperature, while tables in Appendix~\ref{app-a} simply have more data entries.

In astrochemical models, CR-induced physico-chemical surface processes have often been described with the help of a threshold grain temperature, which is reached due to CR impacts and has a certain occurrence frequency (Section~\ref{intro}). For example, \citet{Leger85} and \citet{Shen04} assume 27~K as the triggering temperature for ice chemical explosions, while \citet{Hasegawa93} assume a 70~K threshold for CR-induced evaporation of ice molecules. The 70~K threshold has also been employed to describe CR-induced diffusion and chemical reactions of surface species \citep{Reboussin14,Kalvans14ba,Kalvans15a}. Tables \ref{tab-yr10} and \ref{tab-yr20} display frequencies $\Sigma f_T$ for a range of such minimum $T_{\rm CR}$ values. We also provide the average time, in years, between two successive CR strikes that lift the grain temperature to or above the indicated $T_{\rm CR}$ threshold.

The above data suggest that, in the case of translucent clouds, CR-induced whole-grain heating has a frequency notably higher than that obtained in previous studies. For example, a 0.1~$\mu$m olivine grain is heated to temperatures in excess of 70~K with a frequency $\Sigma f_{>70}=3\times10^{-10}$~s$^{-1}$, compared to $\approx10^{-12}$~s$^{-1}$ assumed previously \citep{Bringa04,Roberts07}. Ice covered grains have $\Sigma f_{>70}$ around $10^{-10}$~s$^{-1}$. The difference spans two orders of magnitude and thus is significant also when the uncertainties of the research are considered. Note that, for grains with $a=0.1$~$\mu$m, $\Sigma f_{>70}$ corresponds to an average temperature of 76-78~K, not 70~K.

Even if $\Sigma f_{>70}$ is ten times lower than our results indicate (because of calculation uncertainties, shielding by magnetic fields, or a higher $N_H$ in starless cloud cores), such a result is important in astrochemistry, for example, due to more efficient CR-induced desorption. Cosmic-ray induced whole-grain heating as a desorption mechanism is used in many independent contemporary astrochemical models \citep[e.g.,][]{Kalvans10,Semenov10,Taquet12,Tassis12,Garrod13,Vasyunin13}. 

A second aspect of high importance is the frequent heating of 10~K grains to or above modest temperatures in the range 20--30~K. Depending on grain size and ice thickness, such heating occurs with $t_{\rm CR}$ intervals of 1-10 years (Table~\ref{tab-yr10}). Given their high frequency, such temperatures are sufficient to overcome energy barriers of up to $\approx800$~K and promote the mobility of atomic H and other light species in the icy mantles of interstellar grains. This allows the synthesis for some species, reducing the radical content in ices.

The short intervals between grain heating events to 20--30~K place constraints on the possibility of chemical explosions in ices \citep{Greenberg73,dHendecourt82,Schutte91}. Such a modest heating, often caused by CR protons, should not induce explosions, otherwise excessive molecule desorption from grains would prevent the accumulation of icy mantles \citep{Ivlev15r}. Despite this, it has been shown that 27~K is a characteristic ignition temperature for irradiated interstellar ice analogs \citep{dHendecourt82}. The radicals necessary for the explosions become mobile at this temperature and are consumed in chemical reactions. Thus, only relatively immobile radicals with high binding energies (OH, NH$_2$) can accumulate. Even their accumulation might not be possible, if grain heating to 20--30~K induces reactions involving abundant ice species. An example is the OH~+~CO reaction with an energy barrier of only 80~K and sufficient surface mobility for CO (binding energy $\approx600$~K).

We conclude that the present results do not favor the possibility of chemical explosions in interstellar ices. Other CR-induced desorption mechanisms, such as spot heating \citep{deJong73,Leger85,Willacy93,Ivlev15r}, sputtering \citep{Draine79}, and ejection of ice fragments \citep{Johnson91,Duley96} probably are more effective than currently assumed in the case of translucent clouds.

Interesting is the dependence of $f_T$ on grain sizes for equal temperatures. The larger grains have larger cross-sections, which means that they receive more CR particle hits. On the other hand, many of these impacts are not effective at significantly raising the grain temperature because of a higher heat capacity. Smaller grains receive less hits, but these are more efficient at raising the grain temperature. Moreover, light element CRs passing through the icy mantle and delivering little energy are sometimes unable to raise the grain temperature even 0.01~K above $T_0$, the resolution limit of this study. In the Appendix tables, they are not accounted for.

\section{Conclusion}
\label{concl}
Summarizing, the paper presents a detailed calculation on whole-grain heating of interstellar grains by CRs at a visual extinction $A_V=2$~mag, corresponding to a translucent molecular cloud. Differences of our calculations and previous studies \citep{Leger85,Shen04,Roberts07} arise primarily because an updated CR energy spectrum was employed (Section~\ref{cr}). The results show that, in the case of translucent clouds, CR-induced whole-grain heating events are about two orders of magnitude more frequent than assumed before. We emphasize the potential significance of CR-induced low-temperature grain heating to influence processes in interstellar ices.

The main results of the present study are datasets for uses in astrochemical numerical simulations. Data in Tables \ref{tab-yr10} and \ref{tab-yr20} can be readily used to describe CR-induced processes in interstellar ices with such models. Moreover, in Appendices \ref{app-a} and \ref{app-b} we report the full $T_{\rm CR}$ and $E_{\rm grain}$ spectra for grains of different sizes and ice layer thickness. These data can be used for more specific investigations considering the interaction between CRs and interstellar grains.

\acknowledgments
I thank Meldra Kemere for the inspiration and initial discussions on this work. This research was supported by the Ventspils City Council and has made use of NASA’s Astrophysics Data System. Finally, I am grateful to the anonymous referee, who helped to improve this article.

\bibliography{cr-dust2}
\bibliographystyle{apj}

\input{CRdust22app}

\end{document}

%% file: CRdust22app.tex
\appendix
\section{Appendix:\\
$T_{\rm CR}$ spectra for grains with an equilibrium temperature $T_0=10$~K}
\label{app-a}

%
\clearpage
\begin{deluxetable}{lccccccc}
\tablewidth{0pt}
\tabletypesize{\scriptsize}
\tablecaption{Calculated temperature frequency $f_T$ spectrum for CR-induced whole-grain heating of 0.05~$\mu$m bare olivine grains with equilibrium temperature of 10~K. Parameters $T_{\rm CR}$ and $E_{\rm grain}$ are the average values for the indicated temperature interval and the corresponding energy interval with a width of $\Delta T_{\rm CR}$ and $\Delta E_{\rm grain}$, respectively.}
\label{tab-A1}
\tablehead{
\colhead{No.} & \colhead{$T_{\rm CR}$, K} & \colhead{interval\tablenotemark{a}, K} & \colhead{$\Delta T_{\rm CR}$, K} & \colhead{$E_{\rm grain}$, eV} & \colhead{interval\tablenotemark{b}, eV} & \colhead{$\Delta E_{\rm grain}$, eV} & \colhead{$f_T$, s$^{-1}$}
}
\startdata
1 & 10.72 & 11.14--11.00 & 0.87 & 6.262E+00 & 1.099--8.966 & 7.816E+00 & 9.124E-14 \\
2 & 11.87 & 11.01--12.00 & 1.00 & 1.898E+01 & ...--2.069E+01 & 1.177E+01 & 3.397E-10 \\
3 & 12.53 & 12.01--13.00 & 1.00 & 2.839E+01 & ...--3.573E+01 & 1.505E+01 & 9.615E-10 \\
4 & 13.49 & 13.01--14.00 & 1.00 & 4.470E+01 & ...--5.468E+01 & 1.895E+01 & 1.151E-09 \\
5 & 14.52 & 14.01--15.00 & 1.00 & 6.640E+01 & ...--7.814E+01 & 2.347E+01 & 9.853E-10 \\
6 & 15.49 & 15.01--16.00 & 1.00 & 9.172E+01 & ...--1.068E+02 & 2.866E+01 & 1.052E-09 \\
7 & 16.50 & 16.01--17.00 & 1.00 & 1.236E+02 & ...--1.410E+02 & 3.418E+01 & 8.869E-10 \\
8 & 17.49 & 17.01--18.00 & 1.00 & 1.612E+02 & ...--1.821E+02 & 4.115E+01 & 9.577E-10 \\
\enddata
\tablenotetext{a}{$T_{\rm CR}$ has a predefined precision of 0.01~K.}
\tablenotetext{b}{For entry No. 1, the full range is indicated. For others, the upper boundary of the interval is indicated, while the lower boundary is that of the previous entry. This pertains to $E_{\rm grain}$ intervals also for other tables in Appendix A.}
\end{deluxetable}

%
\clearpage
\begin{deluxetable}{lccccccc}
\tablewidth{0pt}
\tabletypesize{\scriptsize}
\tablecaption{Calculated temperature frequency $f_T$ spectrum for CR-induced whole-grain heating of 0.05~$\mu$m olivine grains with a thin (0.005~$\mu$m) icy mantle at $T_0=10$~K.}
\label{tab-A2}
\tablehead{
\colhead{No.} & \colhead{$T_{\rm CR}$, K} & \colhead{interval, K} & \colhead{$\Delta T_{\rm CR}$, K} & \colhead{$E_{\rm grain}$, eV} & \colhead{interval, eV} & \colhead{$\Delta E_{\rm grain}$, eV} & \colhead{$f_T$, s$^{-1}$}
}
\startdata
1 & 10.68 & 10.04--11.00 & 0.97 & 1.718E+01 & 1.099--25.78 & 2.468E+01 & 1.362E-09 \\
2 & 11.50 & 11.01--12.00 & 1.00 & 4.104E+01 & ...--5.758E+01 & 3.180E+01 & 2.439E-09 \\
3 & 12.48 & 12.01--13.00 & 1.00 & 7.555E+01 & ...--9.634E+01 & 3.876E+01 & 1.774E-09 \\
4 & 13.46 & 13.01--14.00 & 1.00 & 1.174E+02 & ...--1.429E+02 & 4.660E+01 & 1.624E-09 \\
5 & 14.49 & 14.01--15.00 & 1.00 & 1.693E+02 & ...--1.983E+02 & 5.536E+01 & 1.347E-09 \\
6 & 15.46 & 15.01--16.00 & 1.00 & 2.273E+02 & ...--2.634E+02 & 6.510E+01 & 1.057E-09 \\
7 & 16.48 & 16.01--17.00 & 1.00 & 2.986E+02 & ...--3.384E+02 & 7.504E+01 & 1.046E-09 \\
8 & 17.50 & 17.01--18.00 & 1.00 & 3.817E+02 & ...--4.260E+02 & 8.754E+01 & 8.071E-10 \\
\enddata
\end{deluxetable}

%
\clearpage
\begin{deluxetable}{lccccccc}
\tablewidth{0pt}
\tabletypesize{\scriptsize}
\tablecaption{Calculated temperature frequency $f_T$ spectrum for CR-induced whole-grain heating of 0.05~$\mu$m olivine grains with a thick (0.015~$\mu$m) icy mantle at $T_0=10$~K.}
\label{tab-A3}
\tablehead{
\colhead{No.} & \colhead{$T_{\rm CR}$, K} & \colhead{interval, K} & \colhead{$\Delta T_{\rm CR}$, K} & \colhead{$E_{\rm grain}$, eV} & \colhead{interval, eV} & \colhead{$\Delta E_{\rm grain}$, eV} & \colhead{$f_T$, s$^{-1}$}
}
\startdata
1 & 10.54 & 10.01--11.00 & 1.00 & 3.643E+01 & 1.099--69.78 & 6.868E+01 & 6.524E-09 \\
2 & 11.45 & 11.01--12.00 & 1.00 & 1.062E+02 & ...--1.541E+02 & 8.432E+01 & 4.271E-09 \\
3 & 12.47 & 12.01--13.00 & 1.00 & 2.001E+02 & ...--2.549E+02 & 1.008E+02 & 2.944E-09 \\
4 & 13.49 & 13.01--14.00 & 1.00 & 3.117E+02 & ...--3.739E+02 & 1.189E+02 & 1.894E-09 \\
5 & 14.45 & 14.01--15.00 & 1.00 & 4.354E+02 & ...--5.127E+02 & 1.388E+02 & 1.518E-09 \\
6 & 15.49 & 15.01--16.00 & 1.00 & 5.894E+02 & ...--6.731E+02 & 1.604E+02 & 1.308E-09 \\
7 & 16.52 & 16.01--17.00 & 1.00 & 7.673E+02 & ...--8.550E+02 & 1.819E+02 & 9.851E-10 \\
8 & 17.49 & 17.01--18.00 & 1.00 & 9.574E+02 & ...--1.064E+03 & 2.089E+02 & 7.535E-10 \\
\enddata
\end{deluxetable}

%
\clearpage
\begin{deluxetable}{lccccccc}
\tablewidth{0pt}
\tabletypesize{\scriptsize}
\tablecaption{Calculated temperature frequency $f_T$ spectrum for CR-induced whole-grain heating of 0.1~$\mu$m bare olivine grains at $T_0=10$~K.}
\label{tab-A4}
\tablehead{
\colhead{No.} & \colhead{$T_{\rm CR}$, K} & \colhead{interval, K} & \colhead{$\Delta T_{\rm CR}$, K} & \colhead{$E_{\rm grain}$, eV} & \colhead{interval, eV} & \colhead{$\Delta E_{\rm grain}$, eV} & \colhead{$f_T$, s$^{-1}$}
}
\startdata
1 & 10.77 & 10.01--11.00 & 1.00 & 5.353E+01 & 1.099--71.73 & 7.063E+01 & 5.148E-09 \\
2 & 11.49 & 11.01--12.00 & 1.00 & 1.159E+02 & ...--1.655E+02 & 9.377E+01 & 8.943E-09 \\
3 & 12.49 & 12.01--13.00 & 1.00 & 2.219E+02 & ...--2.859E+02 & 1.204E+02 & 7.047E-09 \\
4 & 13.48 & 13.01--14.00 & 1.00 & 3.561E+02 & ...--4.374E+02 & 1.516E+02 & 5.826E-09 \\
5 & 14.49 & 14.01--15.00 & 1.00 & 5.272E+02 & ...--6.252E+02 & 1.877E+02 & 4.895E-09 \\
6 & 15.49 & 15.01--16.00 & 1.00 & 7.343E+02 & ...--8.544E+02 & 2.293E+02 & 3.869E-09 \\
7 & 16.53 & 16.01--17.00 & 1.00 & 9.968E+02 & ...--1.128E+03 & 2.735E+02 & 3.651E-09 \\
8 & 17.53 & 17.01--18.00 & 1.00 & 1.302E+03 & ...--1.457E+03 & 3.292E+02 & 2.644E-09 \\
\enddata
\end{deluxetable}

%
\clearpage
\begin{deluxetable}{lccccccc}
\tablewidth{0pt}
\tabletypesize{\scriptsize}
\tablecaption{Calculated temperature frequency $f_T$ spectrum for CR-induced whole-grain heating of 0.1~$\mu$m olivine grains with a thin (0.01~$\mu$m) icy mantle at $T_0=10$~K.}
\label{tab-A5}
\tablehead{
\colhead{No.} & \colhead{$T_{\rm CR}$, K} & \colhead{interval, K} & \colhead{$\Delta T_{\rm CR}$, K} & \colhead{$E_{\rm grain}$, eV} & \colhead{interval, eV} & \colhead{$\Delta E_{\rm grain}$, eV} & \colhead{$f_T$, s$^{-1}$}
}
\startdata
1 & 10.51 & 10.01--11.00 & 1.00 & 1.007E+02 & 1.830--206.3 & 2.044E+02 & 2.320E-08 \\
2 & 11.46 & 11.01--12.00 & 1.00 & 3.197E+02 & ...--4.606E+02 & 2.544E+02 & 1.248E-08 \\
3 & 12.47 & 12.01--13.00 & 1.00 & 6.013E+02 & ...--7.707E+02 & 3.101E+02 & 7.952E-09 \\
4 & 13.47 & 13.01--14.00 & 1.00 & 9.419E+02 & ...--1.143E+03 & 3.728E+02 & 5.426E-09 \\
5 & 14.47 & 14.01--15.00 & 1.00 & 1.345E+03 & ...--1.586E+03 & 4.429E+02 & 4.162E-09 \\
6 & 15.47 & 15.01--16.00 & 1.00 & 1.826E+03 & ...--2.107E+03 & 5.208E+02 & 3.319E-09 \\
7 & 16.50 & 16.01--17.00 & 1.00 & 2.405E+03 & ...--2.708E+03 & 6.003E+02 & 2.691E-09 \\
8 & 17.51 & 17.01--18.00 & 1.00 & 3.066E+03 & ...--3.408E+03 & 7.004E+02 & 2.178E-09 \\
\enddata
\end{deluxetable}

%
\clearpage
\begin{deluxetable}{lccccccc}
\tablewidth{0pt}
\tabletypesize{\scriptsize}
\tablecaption{Calculated temperature frequency $f_T$ spectrum for CR-induced whole-grain heating of 0.1~$\mu$m olivine grains with a thick (0.03~$\mu$m) icy mantle at $T_0=10$~K.}
\label{tab-A6}
\tablehead{
\colhead{No.} & \colhead{$T_{\rm CR}$, K} & \colhead{interval, K} & \colhead{$\Delta T_{\rm CR}$, K} & \colhead{$E_{\rm grain}$, eV} & \colhead{interval, eV} & \colhead{$\Delta E_{\rm grain}$, eV} & \colhead{$f_T$, s$^{-1}$}
}
\startdata
1 & 10.37 & 10.01--11.00 & 1.00 & 1.986E+02 & 5.006--558.2 & 5.532E+02 & 5.541E-08 \\
2 & 11.44 & 11.01--12.00 & 1.00 & 8.457E+02 & ...--1.233E+03 & 6.746E+02 & 1.585E-08 \\
3 & 12.46 & 12.01--13.00 & 1.00 & 1.594E+03 & ...--2.039E+03 & 8.064E+02 & 8.094E-09 \\
4 & 13.47 & 13.01--14.00 & 1.00 & 2.477E+03 & ...--2.991E+03 & 9.516E+02 & 5.442E-09 \\
5 & 14.48 & 14.01--15.00 & 1.00 & 3.513E+03 & ...--4.101E+03 & 1.111E+03 & 3.683E-09 \\
6 & 15.49 & 15.01--16.00 & 1.00 & 4.716E+03 & ...--5.385E+03 & 1.284E+03 & 2.769E-09 \\
7 & 16.48 & 16.01--17.00 & 1.00 & 6.082E+03 & ...--6.840E+03 & 1.455E+03 & 2.079E-09 \\
8 & 17.49 & 17.01--18.00 & 1.00 & 7.661E+03 & ...--8.512E+03 & 1.671E+03 & 1.718E-09 \\
\enddata
\end{deluxetable}

%
\clearpage
\begin{deluxetable}{lccccccc}
\tablewidth{0pt}
\tabletypesize{\scriptsize}
\tablecaption{Calculated temperature frequency $f_T$ spectrum for CR-induced whole-grain heating of 0.2~$\mu$m bare olivine grains at $T_0=10$~K.}
\label{tab-A7}
\tablehead{
\colhead{No.} & \colhead{$T_{\rm CR}$, K} & \colhead{interval, K} & \colhead{$\Delta T_{\rm CR}$, K} & \colhead{$E_{\rm grain}$, eV} & \colhead{interval, eV} & \colhead{$\Delta E_{\rm grain}$, eV} & \colhead{$f_T$, s$^{-1}$}
}
\startdata
1 & 10.31 & 10.01--10.50 & 0.50 & 1.647E+02 & 4.925--267.8 & 2.629E+02 & 4.439E-08 \\
2 & 10.74 & 10.51--11.00 & 0.50 & 4.103E+02 & ...--5.738E+02 & 3.060E+02 & 3.706E-08 \\
3 & 11.24 & 11.01--11.50 & 0.50 & 7.366E+02 & ...--9.245E+02 & 3.507E+02 & 2.641E-08 \\
4 & 11.74 & 11.51--12.00 & 0.50 & 1.113E+03 & ...--1.324E+03 & 3.995E+02 & 1.871E-08 \\
5 & 12.25 & 12.01--12.50 & 0.50 & 1.548E+03 & ...--1.777E+03 & 4.526E+02 & 1.595E-08 \\
6 & 12.75 & 12.51--13.00 & 0.50 & 2.025E+03 & ...--2.287E+03 & 5.103E+02 & 1.082E-08 \\
7 & 13.26 & 13.01--13.50 & 0.50 & 2.579E+03 & ...--2.860E+03 & 5.726E+02 & 1.183E-08 \\
8 & 13.80 & 13.51--14.00 & 0.50 & 3.234E+03 & ...--3.499E+03 & 6.399E+02 & 8.699E-09 \\
\enddata
\end{deluxetable}

%
\clearpage
\begin{deluxetable}{lccccccc}
\tablewidth{0pt}
\tabletypesize{\scriptsize}
\tablecaption{Calculated temperature frequency $f_T$ spectrum for CR-induced whole-grain heating of 0.2~$\mu$m olivine grains with a thin (0.02~$\mu$m) icy mantle at $T_0=10$~K.}
\label{tab-A8}
\tablehead{
\colhead{No.} & \colhead{$T_{\rm CR}$, K} & \colhead{interval, K} & \colhead{$\Delta T_{\rm CR}$, K} & \colhead{$E_{\rm grain}$, eV} & \colhead{interval, eV} & \colhead{$\Delta E_{\rm grain}$, eV} & \colhead{$f_T$, s$^{-1}$}
}
\startdata
1 & 10.21 & 10.01--10.50 & 0.50 & 3.254E+02 & 14.64--78.29 & 7.683E+02 & 1.298E-07 \\
2 & 10.73 & 10.51--11.00 & 0.50 & 1.171E+03 & ...--1.650E+03 & 8.672E+02 & 4.509E-08 \\
3 & 11.24 & 11.01--11.50 & 0.50 & 2.098E+03 & ...--2.615E+03 & 9.652E+02 & 2.535E-08 \\
4 & 11.73 & 11.51--12.00 & 0.50 & 3.096E+03 & ...--3.685E+03 & 1.070E+03 & 1.623E-08 \\
5 & 12.23 & 12.01--12.50 & 0.50 & 4.219E+03 & ...--4.866E+03 & 1.181E+03 & 1.309E-08 \\
6 & 12.74 & 12.51--13.00 & 0.50 & 5.486E+03 & ...--6.166E+03 & 1.299E+03 & 9.449E-09 \\
7 & 13.24 & 13.01--13.50 & 0.50 & 6.827E+03 & ...--7.590E+03 & 1.425E+03 & 7.661E-09 \\
8 & 13.75 & 13.51--14.00 & 0.50 & 8.352E+03 & ...--9.148E+03 & 1.558E+03 & 6.072E-09 \\
\enddata
\end{deluxetable}

%
\clearpage
\begin{deluxetable}{lccccccc}
\tablewidth{0pt}
\tabletypesize{\scriptsize}
\tablecaption{Calculated temperature frequency $f_T$ spectrum for CR-induced whole-grain heating of 0.2~$\mu$m olivine grains with a thick (0.06~$\mu$m) icy mantle at $T_0=10$~K.}
\label{tab-A9}
\tablehead{
\colhead{No.} & \colhead{$T_{\rm CR}$, K} & \colhead{interval, K} & \colhead{$\Delta T_{\rm CR}$, K} & \colhead{$E_{\rm grain}$, eV} & \colhead{interval, eV} & \colhead{$\Delta E_{\rm grain}$, eV} & \colhead{$f_T$, s$^{-1}$}
}
\startdata
1 & 10.15 & 10.01--10.50 & 0.50 & 6.373E+02 & 40.05--2131 & 2.090E+03 & 2.702E-07 \\
2 & 10.71 & 10.51--11.00 & 0.50 & 3.117E+03 & ...--4.466E+03 & 2.313E+03 & 4.812E-08 \\
3 & 11.22 & 11.01--11.50 & 0.50 & 5.581E+03 & ...--7.039E+03 & 2.573E+03 & 2.538E-08 \\
4 & 11.74 & 11.51--12.00 & 0.50 & 8.355E+03 & ...--9.862E+03 & 2.824E+03 & 1.533E-08 \\
5 & 12.24 & 12.01--12.50 & 0.50 & 1.131E+04 & ...--1.295E+04 & 3.087E+03 & 1.002E-08 \\
6 & 12.73 & 12.51--13.00 & 0.50 & 1.447E+04 & ...--1.631E+04 & 3.364E+03 & 7.581E-09 \\
7 & 13.23 & 13.01--13.50 & 0.50 & 1.800E+04 & ...--1.997E+04 & 3.654E+03 & 6.137E-09 \\
8 & 13.74 & 13.51--14.00 & 0.50 & 2.186E+04 & ...--2.393E+04 & 3.958E+03 & 4.770E-09 \\
\enddata
\end{deluxetable}

\section{Appendix:\\
$T_{\rm CR}$ spectra for grains with an equilibrium temperature $T_0=20$~K}
\label{app-b}

%
\clearpage
\begin{deluxetable}{lcccc}
\tablewidth{0pt}
\tabletypesize{\scriptsize}
\tablecaption{Calculated temperature frequency $f_T$ spectrum for CR-induced whole-grain heating of 0.05~$\mu$m bare olivine grains at $T_0=20$~K.}
\label{tab-B1}
\tablehead{
\colhead{No.} & \colhead{$T_{\rm CR}$, K} & \colhead{interval, K} & \colhead{$\Delta T_{\rm CR}$, K} & \colhead{$f_T$, s$^{-1}$}
}
\startdata
1 & 20.60 & 20.01--21.00 & 1.00 & 2.944E-09 \\
2 & 21.47 & 21.01--22.00 & 1.00 & 2.563E-09 \\
3 & 22.50 & 22.01--23.00 & 1.00 & 1.635E-09 \\
4 & 23.48 & 23.01--24.00 & 1.00 & 1.123E-09 \\
5 & 24.50 & 24.01--25.00 & 1.00 & 1.017E-09 \\
6 & 25.53 & 25.01--26.00 & 1.00 & 6.828E-10 \\
7 & 26.51 & 26.01--27.00 & 1.00 & 6.623E-10 \\
8 & 27.46 & 27.01--28.00 & 1.00 & 4.519E-10 \\
\enddata
\end{deluxetable}

%
\clearpage
\begin{deluxetable}{lcccc}
\tablewidth{0pt}
\tabletypesize{\scriptsize}
\tablecaption{Calculated temperature frequency $f_T$ spectrum for CR-induced whole-grain heating of 0.05~$\mu$m olivine grains with a thin (0.005~$\mu$m) icy mantle at $T_0=20$~K.}
\label{tab-B2}
\tablehead{
\colhead{No.} & \colhead{$T_{\rm CR}$, K} & \colhead{interval, K} & \colhead{$\Delta T_{\rm CR}$, K} & \colhead{$f_T$, s$^{-1}$}
}
\startdata
1 & 20.45 & 20.01--21.00 & 1.00 & 6.821E-09 \\
2 & 21.45 & 21.01--22.00 & 1.00 & 3.078E-09 \\
3 & 22.49 & 22.01--23.00 & 1.00 & 1.745E-09 \\
4 & 23.52 & 23.01--24.00 & 1.00 & 1.147E-09 \\
5 & 24.50 & 24.01--25.00 & 1.00 & 8.344E-10 \\
6 & 25.47 & 25.01--26.00 & 1.00 & 6.612E-10 \\
7 & 26.50 & 26.01--27.00 & 1.00 & 5.704E-10 \\
8 & 27.49 & 27.01--28.00 & 1.00 & 4.136E-10 \\
\enddata
\end{deluxetable}

%
\clearpage
\begin{deluxetable}{lcccc}
\tablewidth{0pt}
\tabletypesize{\scriptsize}
\tablecaption{Calculated temperature frequency $f_T$ spectrum for CR-induced whole-grain heating of 0.05~$\mu$m olivine grains with a thick (0.015~$\mu$m) icy mantle at $T_0=20$~K.}
\label{tab-B3}
\tablehead{
\colhead{No.} & \colhead{$T_{\rm CR}$, K} & \colhead{interval, K} & \colhead{$\Delta T_{\rm CR}$, K} & \colhead{$f_T$, s$^{-1}$}
}
\startdata
1 & 20.35 & 20.01--21.00 & 1.00 & 1.444E-08 \\
2 & 21.43 & 21.01--22.00 & 1.00 & 3.673E-09 \\
3 & 22.47 & 22.01--23.00 & 1.00 & 1.903E-09 \\
4 & 23.50 & 23.01--24.00 & 1.00 & 1.164E-09 \\
5 & 24.48 & 24.01--25.00 & 1.00 & 7.876E-10 \\
6 & 25.48 & 25.01--26.00 & 1.00 & 6.271E-10 \\
7 & 26.48 & 26.01--27.00 & 1.00 & 4.629E-10 \\
8 & 27.47 & 27.01--28.00 & 1.00 & 3.944E-10 \\
\enddata
\end{deluxetable}

%
\clearpage
\begin{deluxetable}{lcccc}
\tablewidth{0pt}
\tabletypesize{\scriptsize}
\tablecaption{Calculated temperature frequency $f_T$ spectrum for CR-induced whole-grain heating of 0.1~$\mu$m bare olivine grains at $T_0=20$~K.}
\label{tab-B4}
\tablehead{
\colhead{No.} & \colhead{$T_{\rm CR}$, K} & \colhead{interval, K} & \colhead{$\Delta T_{\rm CR}$, K} & \colhead{$f_T$, s$^{-1}$}
}
\startdata
1 & 20.41 & 20.01--21.00 & 1.00 & 3.000E-08 \\
2 & 21.47 & 21.01--22.00 & 1.00 & 9.484E-09 \\
3 & 22.47 & 22.01--23.00 & 1.00 & 4.865E-09 \\
4 & 23.46 & 23.01--24.00 & 1.00 & 3.479E-09 \\
5 & 24.45 & 24.01--25.00 & 1.00 & 2.310E-09 \\
6 & 25.47 & 25.01--26.00 & 1.00 & 1.878E-09 \\
7 & 26.48 & 26.01--27.00 & 1.00 & 1.344E-09 \\
8 & 27.46 & 27.01--28.00 & 1.00 & 1.062E-09 \\
\enddata
\end{deluxetable}

%
\clearpage
\begin{deluxetable}{lcccc}
\tablewidth{0pt}
\tabletypesize{\scriptsize}
\tablecaption{Calculated temperature frequency $f_T$ spectrum for CR-induced whole-grain heating of 0.1~$\mu$m olivine grains with a thin (0.01~$\mu$m) icy mantle at $T_0=20$~K.}
\label{tab-B5}
\tablehead{
\colhead{No.} & \colhead{$T_{\rm CR}$, K} & \colhead{interval, K} & \colhead{$\Delta T_{\rm CR}$, K} & \colhead{$f_T$, s$^{-1}$}
}
\startdata
1 & 20.30 & 20.01--21.00 & 1.00 & 4.791E-08 \\
2 & 21.43 & 21.01--22.00 & 1.00 & 9.203E-09 \\
3 & 22.46 & 22.01--23.00 & 1.00 & 4.512E-09 \\
4 & 23.47 & 23.01--24.00 & 1.00 & 2.707E-09 \\
5 & 24.46 & 24.01--25.00 & 1.00 & 1.886E-09 \\
6 & 25.48 & 25.01--26.00 & 1.00 & 1.390E-09 \\
7 & 26.48 & 26.01--27.00 & 1.00 & 1.102E-09 \\
8 & 27.49 & 27.01--28.00 & 1.00 & 8.717E-10 \\
\enddata
\end{deluxetable}

%
\clearpage
\begin{deluxetable}{lcccc}
\tablewidth{0pt}
\tabletypesize{\scriptsize}
\tablecaption{Calculated temperature frequency $f_T$ spectrum for CR-induced whole-grain heating of 0.1~$\mu$m olivine grains with a thick (0.03~$\mu$m) icy mantle at $T_0=20$~K.}
\label{tab-B6}
\tablehead{
\colhead{No.} & \colhead{$T_{\rm CR}$, K} & \colhead{interval, K} & \colhead{$\Delta T_{\rm CR}$, K} & \colhead{$f_T$, s$^{-1}$}
}
\startdata
1 & 20.22 & 20.01--21.00 & 1.00 & 8.045E-08 \\
2 & 21.42 & 21.01--22.00 & 1.00 & 8.884E-09 \\
3 & 22.46 & 22.01--23.00 & 1.00 & 3.927E-09 \\
4 & 23.47 & 23.01--24.00 & 1.00 & 2.298E-09 \\
5 & 24.47 & 24.01--25.00 & 1.00 & 1.703E-09 \\
6 & 25.49 & 25.01--26.00 & 1.00 & 1.245E-09 \\
7 & 26.49 & 26.01--27.00 & 1.00 & 9.992E-10 \\
8 & 27.50 & 27.01--28.00 & 1.00 & 8.185E-10 \\
\enddata
\end{deluxetable}

%
\clearpage
\begin{deluxetable}{lcccc}
\tablewidth{0pt}
\tabletypesize{\scriptsize}
\tablecaption{Calculated temperature frequency $f_T$ spectrum for CR-induced whole-grain heating of 0.2~$\mu$m bare olivine grains at $T_0=20$~K.}
\label{tab-B7}
\tablehead{
\colhead{No.} & \colhead{$T_{\rm CR}$, K} & \colhead{interval, K} & \colhead{$\Delta T_{\rm CR}$, K} & \colhead{$f_T$, s$^{-1}$}
}
\startdata
1 & 20.16 & 20.01--20.50 & 0.50 & 1.486E-07 \\
2 & 20.71 & 20.51--21.00 & 0.50 & 3.214E-08 \\
3 & 21.23 & 21.01--21.50 & 0.50 & 1.590E-08 \\
4 & 21.73 & 21.51--22.00 & 0.50 & 9.555E-09 \\
5 & 22.25 & 22.01--22.50 & 0.50 & 7.017E-09 \\
6 & 22.74 & 22.51--23.00 & 0.50 & 4.056E-09 \\
7 & 23.23 & 23.01--23.50 & 0.50 & 4.304E-09 \\
8 & 23.75 & 23.51--24.00 & 0.50 & 2.843E-09 \\
\enddata
\end{deluxetable}

%
\clearpage
\begin{deluxetable}{lcccc}
\tablewidth{0pt}
\tabletypesize{\scriptsize}
\tablecaption{Calculated temperature frequency $f_T$ spectrum for CR-induced whole-grain heating of 0.2~$\mu$m olivine grains with a thin (0.02~$\mu$m) icy mantle at $T_0=20$~K.}
\label{tab-B8}
\tablehead{
\colhead{No.} & \colhead{$T_{\rm CR}$, K} & \colhead{interval, K} & \colhead{$\Delta T_{\rm CR}$, K} & \colhead{$f_T$, s$^{-1}$}
}
\startdata
1 & 20.13 & 20.01--20.50 & 0.50 & 2.128E-07 \\
2 & 20.71 & 20.51--21.00 & 0.50 & 2.774E-08 \\
3 & 21.23 & 21.01--21.50 & 0.50 & 1.280E-08 \\
4 & 21.74 & 21.51--22.00 & 0.50 & 7.557E-09 \\
5 & 22.25 & 22.01--22.50 & 0.50 & 5.151E-09 \\
6 & 22.75 & 22.51--23.00 & 0.50 & 3.756E-09 \\
7 & 23.24 & 23.01--23.50 & 0.50 & 3.040E-09 \\
8 & 23.74 & 23.51--24.00 & 0.50 & 2.463E-09 \\
\enddata
\end{deluxetable}

%
\clearpage
\begin{deluxetable}{lcccc}
\tablewidth{0pt}
\tabletypesize{\scriptsize}
\tablecaption{Calculated temperature frequency $f_T$ spectrum for CR-induced whole-grain heating of 0.2~$\mu$m olivine grains with a thick (0.06~$\mu$m) icy mantle at $T_0=20$~K.}
\label{tab-B9}
\tablehead{
\colhead{No.} & \colhead{$T_{\rm CR}$, K} & \colhead{interval, K} & \colhead{$\Delta T_{\rm CR}$, K} & \colhead{$f_T$, s$^{-1}$}
}
\startdata
1 & 20.10 & 20.01--20.50 & 0.50 & 2.957E-07 \\
2 & 20.71 & 20.51--21.00 & 0.50 & 2.455E-08 \\
3 & 21.23 & 21.01--21.50 & 0.50 & 1.118E-08 \\
4 & 21.75 & 21.51--22.00 & 0.50 & 6.823E-09 \\
5 & 22.25 & 22.01--22.50 & 0.50 & 4.771E-09 \\
6 & 22.76 & 22.51--23.00 & 0.50 & 3.533E-09 \\
7 & 23.25 & 23.01--23.50 & 0.50 & 2.874E-09 \\
8 & 23.75 & 23.51--24.00 & 0.50 & 2.468E-09 \\
\enddata
\end{deluxetable}